\documentclass[aps,prd,superscriptaddress,groupedaddress,nofootinbib,longbibliography,preprint,floatfix]{revtex4-1}
\newcommand{\bew}{\begin{widetext}}
\newcommand{\enw}{\end{widetext}}
\newcommand{\bee}{\begin{equation}}
\newcommand{\ene}{\end{equation}}
\newcommand{\bea}{\begin{eqnarray}}
\newcommand{\ena}{\end{eqnarray}}
\newcommand{\beal}{\begin{align}}
\newcommand{\enal}{\end{align}}
\newcommand{\beald}{\begin{aligned}}
\newcommand{\enald}{\end{aligned}}
\newcommand{\bes}{\begin{subequations}}
\newcommand{\ens}{\end{subequations}}

\def\to{\rightarrow}

\def\met{E_T \hspace*{-1.1em}/\hspace*{0.5em}}

\def\mev{\,{\rm MeV}}
\def\gev{\,{\rm GeV}}
\def\tev{\,{\rm TeV}}

\def\fb{\,{\rm fb}}

\def\ifb{$\,{\rm fb}^{-1}$}

\def\iab{$\,{\rm ab}^{-1}$}

\usepackage{slashed}
\usepackage{multirow}
\usepackage{longtable}
\usepackage{amsmath,amssymb,amsfonts}
\usepackage[colorlinks]{hyperref}
\usepackage{graphicx}
\usepackage{subfigure}
\graphicspath{{./Nudamol/}}
\allowdisplaybreaks

\def\to{\rightarrow}

\def\tev{{\rm TeV}}
\def\gev{{\rm GeV}}

\begin{document}
%\linenumbers

%%%%%%%%%%%%%%%%%%%%%%%%%%%%%%%%%%%%%%%%%%%%%

\title{Mono-lepton Signature of a Neutrino-philic Dark Fermion at Hadron Colliders}

\author{Kai Ma}
\email{kai@xauat.edu.cn}
\affiliation{Faculty of Science, Xi'an University of Architecture and Technology, Xi'an, 710055, China}

\author{Lin-Yun He}
\email{a1164432527@gmail.com}
\affiliation{Center of Advanced Quantum Studies, School of Physics and Astronomy, 
Beijing Normal University, Beijing, 100875, China}
\affiliation{Faculty of Science, Xi'an University of Architecture and Technology, Xi'an, 710055, China}

\date{\today}

\begin{abstract}
Searching for dark matter at high-energy colliders and direct detection experiments can effectively cover nearly the entire mass range from the MeV to the TeV scale. In this paper, we focus on four-fermion contact interactions formulated within the framework of Effective Field Theory. Specifically, we present a detailed analysis of mono-lepton production at the LHC. Our results demonstrate that tensor operators exhibit superior sensitivity in the mono-lepton channel, constraining energy scales up to 3 TeV for a nearly massless dark fermion using current LHC data. Moreover, these operators mediate both spin-independent and spin-dependent absorption processes in nuclear targets. A systematic comparison of constraints between direct detection experiments and collider measurements reveals the LHC's distinct advantage in exploring sub-GeV dark matter candidates while maintaining competitive sensitivity at the TeV scale. Notably, direct detection experiments such as Super-Kamiokande and Borexino achieve complementary constraints in the 10-100 MeV mass range through their unique capabilities: utilization of light nuclei targets, large exposure volumes, and distinctive features of the recoil energy spectra.
\end{abstract}

\maketitle
\setcounter{page}{1}
\renewcommand{\thefootnote}{\arabic{footnote}}
\setcounter{footnote}{0}
\tableofcontents

\section{Introduction}
\label{sec:intro}
For several decades, dark matter (DM) has been acknowledged as an essential and fundamental component of the universe. This recognition is firmly supported by compelling and robust evidence derived from various aspects of galactic dynamics as well as comprehensive cosmological observations, as detailed in references \cite{Navarro:1995iw,Clowe:2006eq,Spergel:1999mh,Bertone:2004pz}.
Despite the extensive and intensive efforts in conducting numerous searches,
both within the high energy collider searches  \cite{Boveia:2018yeb,Arcadi:2017atc,Landsberg:2015kwa} and through astrophysical measurements \cite{Steigman:2012nb,deDiosZornoza:2021rgw,PerezdelosHeros:2020qyt}, no definitive and conclusive signals of DM have been detected.
As a direct consequence, the intrinsic properties of DM remain largely enigmatic and undetermined. This lack of knowledge has spurred the community to continuously explore new experimental techniques and theoretical frameworks in an attempt to unlock the mysteries of DM.

Recent developments have brought the absorption operator of fermionic DM into sharper focus. This operator is intended to shed light on the interactions between the DM and the Standard Model (SM) particles \cite{Cox:2023cjw,Ge:2024euk,Ma:2024tkt,Ma:2024aoc},
and can be explored in direct detection experiments with various target,
such as semiconductors \cite{Hochberg:2016sqx},
superconductors \cite{Hochberg:2016ajh},
phonons \cite{Mitridate:2023izi},
 electrons \cite{Ge:2022ius,PandaX:2022ood}.
The CDEX Collaboration has recently set upper limits on the absorption
cross-section of DM with a mass of 5 ${\rm keV}$
when it interacts with electron targets \cite{CDEX:2024bum}.
For vector interactions, the measured upper bound is $5.5 \times 10^{-46}\,{\rm cm^2}$
and for axial-vector interactions, it was $1.8 \times 10^{-46}\, {\rm cm^2}$.
The above result impose crucial constraints on both the theoretical models
being developed and the experimental approaches being employed to detect
and describe the characteristics of DM.

In case of that the DM is stabilised due to an underlying discrete symmetry like $\mathbb{Z}_2$, DM particles are usually produced in pairs in collider experiments.
In contrast, there is no such symmetry in the absorptive operators.
The DM particle can be stable only in case of that it has a low enough mass,
such that its decay width can be minimized,
and allow it to persist until now as a viable DM candidate
with the correct relic abundance \cite{Dror:2020czw,Ge:2022ius,Ge:2023wye}.
Correspondingly, the collider singles are also completely different.
In this paper, we explore the mono-lepton production
which has been recognized as a promising channel in searching for missing energy
\cite{Bai:2012xg,CMS:2013iea,Matsumoto:2018ioi}.
The unique advantage of mono-lepton events lies in their minimal background noise, which significantly enhances the experimental sensitivity to various interaction operators between the dark particle and SM particles.

We look into four-fermion contact interactions within the
Effective Field Theory (EFT) framework.
We are specifically interested in the interactions that involve a fermionic DM,
a neutrino, and two quarks.
These effective operators have already been studied quite a lot in earlier works
\cite{Dreiner:2013vla,Bishara:2017pfq,Belyaev:2018pqr,Ma:2024tkt}.
Our analysis is centered on the mono-lepton process.
where the outgoing DM and neutrino behaves as missing energy.
In addition, we examine the parton-level absorption processes
at nuclear targets that are brought about by these four-fermion interactions.
We also compare the exclusion limits that come from direct detection experiments
and LHC data.
By doing this, we can provide useful predictions for future research directions.

The rest of this paper is organized as follows. In  Sec.~\ref{sec:eft},
we give a short review of the four-fermion contact interactions
and the validity of the EFT framework  in high energy condition.
In Sec.~\ref{sec:LX:LHC}, we study the production properties of both the signal
and the irreducible background of the mono-lepton process.
and the constraints from the (future) hadron colliders are also studied.
In Sec.~\ref{sec:Absor},
we study the absorption process of a light DM on a nuclei target,
both for the spin-independent and spin-dependent scatterings.
The combined constraints from the hadron collider searches and
direct detections are also discussed.
Our conclusions are given in the final section \ref{sec:conclusion}.

\section{Effective Operators}
\label{sec:eft}
At hadron collider, the four-fermion effective operators can be investigated via mono-jet \cite{Ma:2024gqj,Abdallah:2015uba,Bai:2015nfa}, mono-photon \cite{daSilveira:2023hmt,Kalinowski:2022unu,Gershtein:2008bf}, and mono-Z/W processes \cite{H:2024mff,Benitez-Irarrazabal:2024ich,Kundu:2021cmo,Wan:2018eaz,Bell:2015rdw} which can summarized as mono-X processes \cite{Acharya:2024vsu,Bhattacharya:2022qck,Bernreuther:2018nat}.   These four-fermion contact interactions can also lead to the absorption of fermionic DM by nuclei \cite{PandaX:2022osq,Dror:2019dib,Li:2022kca,Li:2022kca}, where the DM mass and energy are transferred to the nuclear recoil energy \cite{Dror:2019onn,Belfatto:2021ats,Dror:2019dib,Ma:2024aoc}.
 In the EFT framework, such lowest
order interactions as encapsulated by the following dimension-6 operators:
\begin{equation}
\label{eq:effo}
\begin{aligned}
\mathcal{O}_{S} & \equiv(\bar{q} q)\left(\bar{\nu}_{L} \chi_{R}\right) \,,
\\
\mathcal{O}_{P} & \equiv\left(\bar{q} i \gamma_{5} q\right)\left(\bar{\nu}_{L} \chi_{R}\right) \,,
\\
\mathcal{O}_{V} & \equiv\left(\bar{q} \gamma_{\mu} q\right)\left(\bar{\nu}_{L} \gamma^{\mu} \chi_{L}\right) \,,
\\
\mathcal{O}_{A} & \equiv\left(\bar{q} \gamma_{\mu} \gamma_{5} q\right)\left(\bar{\nu}_{L} \gamma^{\mu} \chi_{L}\right) \,,
\\
\mathcal{O}_{T} & \equiv\left(\bar{q} \sigma_{\mu \nu} q\right)\left(\bar{\nu}_{L} \sigma^{\mu \nu} \chi_{R}\right) \,,
\end{aligned}
\end{equation}
Here, $q$ denotes an isospin doublet, $q=(u~d)^T$, representing the first generation of quarks. The operators link the quark current with the DM-neutrino current, capturing potential interactions across these fermionic fields. The corresponding effective Lagrangian is expressed as:
\bee
\mathcal{L}_{\text {eff }}
=
\sum_{i} \frac{1}{\varLambda_{i}^{2}} \mathcal{O}_{i}+\text { h.c.. }
\ene
where $\Lambda_i$ denotes the energy scale associated with each operator. It should be noted that the above effective Lagrangian in general should be parameterized by \(c_i/M_i\) instead of \(1/\varLambda_i\), where \(M_i\) is the mass scale of possible new physics and $c_i$ are the corresponding Wilson coefficients at electroweak scale.
In practice, the Wilson coefficients \(c_i\) can be either as small as zero or as large as \(4\pi\) (the upper bound arises from the perturbative validity). 
Given the range of \(c_i\), the energy scale \(\varLambda_i\) can thus be either smaller or larger than the mass scale \(M_i\) of possible new physics\cite{Liu:2016idz,Giudice:2007fh}.
In our case, since the operators under consideration cannot resolve \(c_i/M_i\), the coefficients are universally normalized to 1. This formalism allows for the systematic exploration of beyond SM physics at accessible large energy scales, examining potential signals and constraints from particle physics experiments and astrophysical observations.

To ensure the integrity of the EFT framework outlined above, several conditions must be met regarding the energy scales and the renormalization group effects. Specifically, the energy scale $\Lambda_i$ associated with the EFT operators must exceed the energy of the particle collisions involved \cite{Dreiner:2013vla}. Furthermore, if a mediator is involved, its mass should be significantly greater than the collision energy to avoid invalidating the EFT approximation \cite{Busoni:2013lha,Busoni:2014sya,Busoni:2014haa}. Given that our EFT operators are defined at the $\tev$ scale, renormalization group (RG) effects become particularly significant at lower energy scales, such as those encountered during absorption processes \cite{Hill:2011be,Frandsen:2012db,Vecchi:2013iza,Crivellin:2014qxa,DEramo:2014nmf,DEramo:2016gos,Belyaev:2018pqr}. These RG effects, due to the lack of Galilean invariance, can lead to a mixing of operators with differing Lorentz structures when applied in non-relativistic scenarios \cite{Bishara:2017pfq}. However, the extent of this mixing effect hinges significantly on the strength of the coupling between DM and the Standard Model fermions, particularly the top quark \cite{Hill:2011be,Frandsen:2012db,Vecchi:2013iza,Crivellin:2014qxa,DEramo:2014nmf,DEramo:2016gos,Bishara:2017pfq,Belyaev:2018pqr}.

\begin{figure}[h]
\centering
\includegraphics[width=0.58\textwidth]{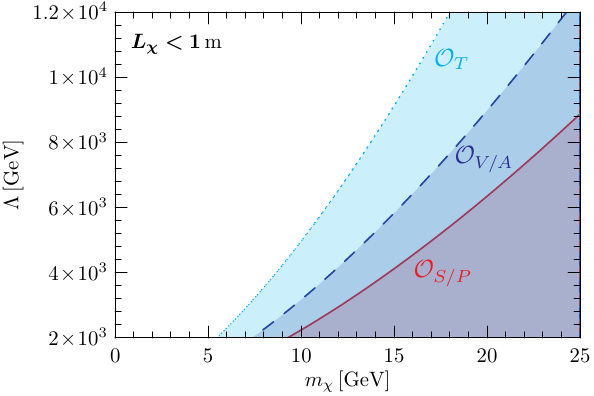}
\caption{\it
Typical decay length of the dark fermion in the $m_\chi$-$\varLambda$ plane.
The parton level center of mass energy is chosen as $\sqrt{\hat{s}}=1\tev$.
}
\label{fig:dcylen}
\end{figure}

Furthermore, the dark fermion can decay only
through following channel,
\bee
\chi \to \nu + q + \overline{q}
\ene
and the corresponding total decay widths are given as,
\begin{equation}
\begin{aligned}
&\Gamma_{S/P}=\frac{m_\chi^5}{1024\pi^3 \Lambda^4_{S/P}} \,,
\\
&\Gamma_{V/A}=\frac{m_\chi^5}{256\pi^3 \Lambda^4_{V/A}} \,,
\\
&\Gamma_{T}=\frac{3m_\chi^5}{128\pi^3 \Lambda^4_{T}} \,,
\end{aligned}
\end{equation}
One can observe that,
in general, the total decay widths for all the operators depend on $\frac{m^5_\chi}{\Lambda^4_i}$ .
If the dark fermion  is very light, specifically when $ m_\chi < \varLambda_{\rm QCD}$, the aforementioned decay channels are prohibited because of the quark pair condensation.
Conversely, signals of the four-fermion contact couplings may show up in the
invisible decay of neutral pion $\pi \rightarrow \nu \bar\chi$ \cite{PIENU:2017wbj}.
Since there is no direct signal, this channel is very difficult to detect.
However, this mass range is outside the range of interest for a light DM
(studied in the Sec. \ref{sec:Absor} for direct detections.
On the other hand, at a collider with (parton level) center-of-mass energy
$\sqrt{\hat{s}}$, the typical decay length of the dark fermion is given as,
\bee
L_{\chi}
= \gamma_{\chi} \tau_{\chi}
= \frac{ \sqrt{\hat{s}} }{2 m_\chi \varGamma_{\chi} }
\Big( 1 + \frac{ m_\chi^2 }{ \sqrt{\hat{s}} } \Big)\,.
\ene
For a relatively light dark fermion, the following relation holds:
$L_{\chi}  \approx  \sqrt{\hat{s}}/(2 m_\chi \varGamma_{\chi} ) $.
As a reference, we simply stipulate that the typical decay length should be greater than 1 m.
The Fig.~\ref{fig:dcylen} illustrates the regions where $L_\chi < 1\,$m
for a typical center-of-mass energy at the LHC, which is $\sqrt{\hat{s}}=1\tev$.
One can observe that for a dark fermion with a mass $m_\chi \lesssim 5\gev$,
the assumption of invisibility is consistently valid.
Moreover, for a dark fermion with $m_\chi \lesssim 5\gev$,
there exist various thermal (and non-thermal) production mechanisms that
can account for the observed DM relic abundance \cite{PIENU:2017wbj}.
In this paper, instead of delving into the details of such production mechanisms,
we conduct a study in a model-independent manner on the signal properties
in hadron collider and direct detection experiments.
For a relatively heavier dark fermion,
the restriction regarding invisibility is dependent on the energy scale $\varLambda_i$.
However, it must be pointed out that this constraint is model-dependent
rather than a general one, especially when the dark fermion $\chi$ is just one part of
an entire dark sector, as referenced in
\cite{Gori:2022vri,Marra:2019lyc,Deliyergiyev:2015oxa,Hofmann:2020wvr,Lagouri:2022ier}.
Taking into account the advantage of high-energy colliders, that is,
their ability to search for much heavier particles,
we will investigate the signals of a dark fermion across the full mass region
as long as it is kinematically feasible.

\section{Mono-lepton Production}
\label{sec:LX:LHC}
\begin{figure}[tbh]
\centering
\includegraphics[height=0.12\textheight]{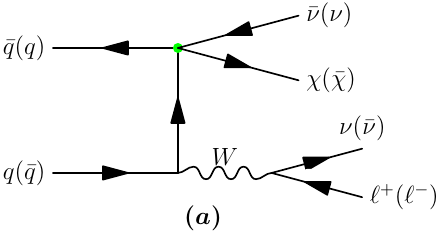}
\quad
\includegraphics[height=0.12\textheight]{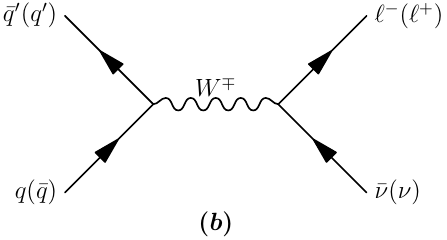}
\caption{\it
Feynman diagrams of the mono-lepton process $pp \to \ell^\pm\slashed{E}_T + X$:
\textbf{(a)} is for the signal operators,
\textbf{(b)} is for the irreducible background.
}
\label{fig:Feyn:Monol}
\end{figure}

As we have mentioned, the mono-lepton production serves as a promising channel
in collider searches of missing energy. In our case the process can be induced
by the four-fermion operators defined in \eqref{eq:effo},
and the parton-level channels are given as,
\bee
q + \bar{q}' \;\to\; \ell^\pm + \nu + \bar\nu + \chi (\bar\chi) \,.
%p + p \;\to\; \ell^\pm + \slashed{E}_T + X\,,
%q + \bar{q}' \;\to\; \ell^\pm + \slashed{E}_T \bar\chi\, \text{and}~ \bar\ell + \chi \,.
\ene
Here the neutrinos and the dark fermion $\chi(\bar\chi)$ beehives as missing energy,
and hence the transverse missing energy ($\slashed{E}_T$) is the natural observable
of looking for the signals.
Fig.~\ref{fig:Feyn:Monol}(a) depicts the parton-level Feynman diagrams for the signal,
while Fig.~\ref{fig:Feyn:Monol}(b) shows the corresponding irreducible background.
The mono-$W^\pm$ production, subsequent to which leptonic decays occur,
gives rise to an irreducible background. This background can be effectively mitigated
by imposing an appropriate kinematic cut on the transverse mass of the lepton
and the transverse missing energy.
\begin{figure}[ptbh]
\centering
\includegraphics[width=0.44\textwidth]{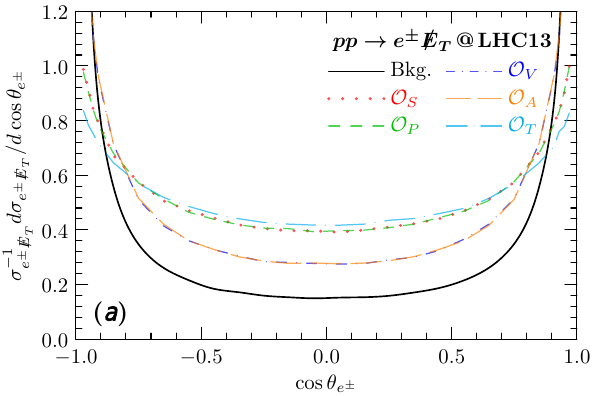}
\quad
\includegraphics[width=0.44\textwidth]{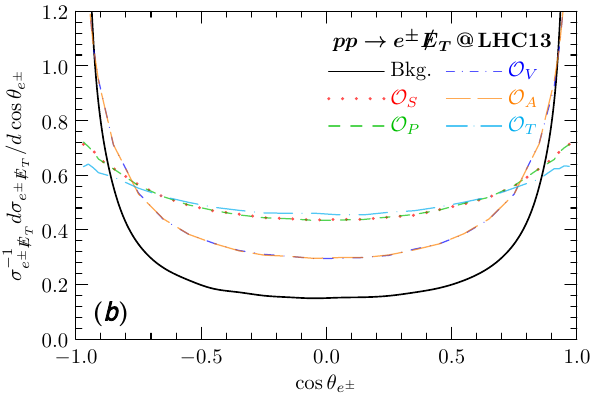}
\caption{\it
The parton-level normalized distributions of the polar angle of the outgoing electron
($\theta_{e^\pm}$) are presented for a center-of-mass (CoM) energy of
$\sqrt{s} = 13\tev$.
The signal events are depicted by colorful curves. In panel \textbf{(a)}, the mass parameter is set to $m_\chi = 0\gev$, and in panel \textbf{(b)},
it is set to $m_\chi = 1\tev$.
The irreducible background is represented by a solid black curve.
}
\label{fig:XCE-YNXS:LHC13}
\end{figure}
Panels (a) and (b) of Fig.~\ref{fig:XCE-YNXS:LHC13} exhibit normalized distributions of the polar angle (\(\theta_{e^\pm}\)). Since the normalized differential cross sections are independent of the energy scale \(\varLambda_i\), we have used a typical value \(\varLambda_i = 1\) TeV in our numerical calculation. While the mass parameter can have a non-trivial effect on the polar angle distribution. As shown in Fig.~\ref{fig:XCE-YNXS:LHC13}(a), the case for \(m_\chi = 0\) GeV is displayed, while Fig.~\ref{fig:XCE-YNXS:LHC13}(b) shows the case for \(m_\chi = 1\) TeV. The irreducible background is delineated by a solid-black curve, and the signal events are characterized by 
by colorful non-solid curves.
It is readily apparent that within the central region,
the discrepancies in the polar angle distribution are insignificant
for differentiating between the signal and the background.
This can be predominantly attributed to the fact that
 $W^\pm$  bosons are emitted along the direction of the incoming partons.
In the forward and backward regions,
both the signal events and the irreducible background are conspicuously intense.
As shown in Fig.~\ref{fig:XCE-YNXS:LHC13}(b),
the divergence among the signal events induced by the four operators
with distinct Lorentz structures becomes more pronounced as $m_\chi$ increases.
However, in practical scenarios,
discerning this divergence may prove to be extremely difficult.
The substantial background in these regions has the potential
to obscure the unique characteristics of signal operators.

\begin{figure}[th]
\centering
\includegraphics[width=0.44\textwidth]{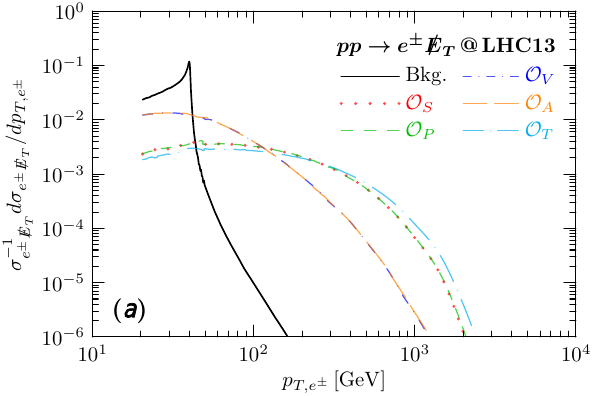}
\quad
\includegraphics[width=0.44\textwidth]{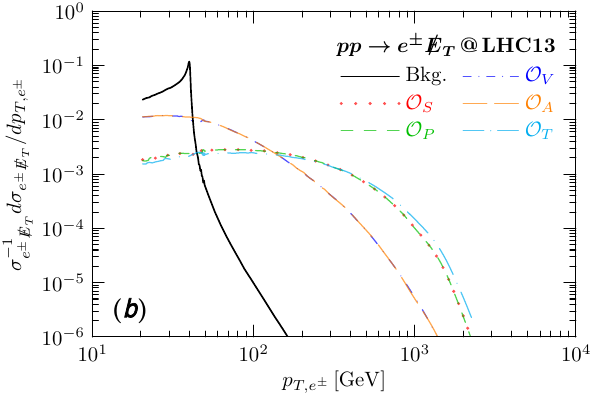}
\caption{\it
The normalized parton-level distributions of the transverse momentum of
the outgoing electron ($p_{T,e^\pm}$) are presented for a center-of-mass energy of
$\sqrt{s} = 13\tev$.
The signal events are represented by colorful lines.
For the panel \textbf{(a)} , the signal corresponds to parameters
$\varLambda_i = 1\tev$ and $m_\chi = 0\gev$, while for the panel \textbf{(b)},
the signal is characterized by $\varLambda_i = 1\tev$ and $m_\chi = 1\tev$.
The irreducible background is depicted by a solid black curve.
}
\label{fig:DXS:LX:LHC13}
\end{figure}
The utility of the mono-lepton process is clearly demonstrated by analyzing
the transverse momentum distributions in Figs.~\ref{fig:DXS:LX:LHC13}(a) and (b).
Charged leptons from $W^{\pm}$ boson decays typically
have a transverse momentum peak at around half the $W^{\pm}$ boson mass,
i.e., $p_{\ell, T} \sim m_W/2$.
This characteristic is crucial for analyzing the signal under different conditions:
$\varLambda_i = 1\tev$, $m_\chi = 0\gev$ in Fig.~\ref{fig:DXS:LX:LHC13}(a)
and $m_\chi = 1\tev$ in Fig.~\ref{fig:DXS:LX:LHC13}(b).
In both cases, the transverse momentum distribution of the irreducible background,
shown as a black solid curve, has a prominent peak and rapidly declines
when $(p_T,e^\pm)$ exceeds the mass threshold.
This feature helps distinguish the signal from the background,
improving the dark-fermion detection sensitivity via the mono-lepton channel.
Moreover, the transverse momentum distributions intuitively show the differences
in Lorentz structures between operators.
Signals induced by $\mathcal{O}_{S}$, $\mathcal{O}_{P}$,
especially $\mathcal{O}_{P}$, dominate the higher $(p_T,e^\pm)$ region.
This suggests that the mono-lepton search is highly sensitive to the tensor operator,
highlighting its effectiveness in detecting high-transverse-momentum signals.

\begin{figure}[th]
\centering
\includegraphics[width=0.48\textwidth]{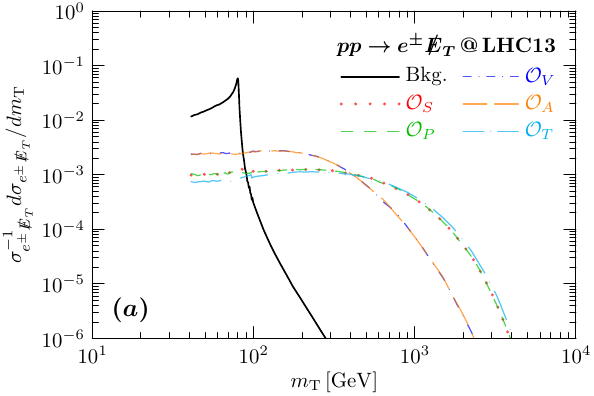}
\quad
\includegraphics[width=0.48\textwidth]{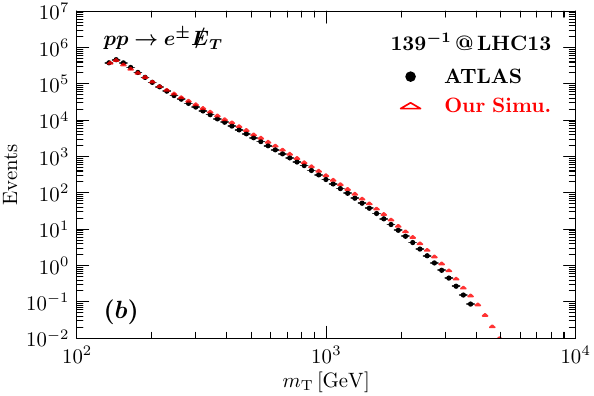}
\caption{\it
\textbf{Panel (a)}: The normalized parton-level distributions of the transverse mass ($m_T$) of the outgoing electron are presented for a center-of-mass (CoM) energy of $\sqrt{s} = 13\tev$. The signal events with $m_\chi = 0\gev$ and $\varLambda_i = 1\tev$ are represented by colorful non-solid curves, while the irreducible background is represented by a solid black curve.
\textbf{Panel (b)}: Validation of our simulation for the transverse mass ($m_T$) distribution of the irreducible background channel $pp\to W^\pm \to e^\pm \nu$ at the LHC is performed with a total luminosity of $\mathcal{L} = 139\fb^{-1}$. The experimental data (black dots) are taken from Ref.~\cite{ATLAS:2021kxv}, and our results (red rectangles) have been renormalized by multiplying an overall constant.
}
\label{fig:LX:MT:LHC13}
\end{figure}
Fig.~\ref{fig:LX:MT:LHC13}(a) presents the normalized parton-level distribution of the transverse mass. A highly conspicuous peak at $m_T = m_W$ is discernible in the background, which is rigorously defined by the formula:
\bee
m_T = \sqrt{ 2 p_{\ell, T} \slashed{E}_T (1 - \cos\phi_{\ell\nu} ) } \,.
\ene
In this context, $p_{\ell, T}$ stands for the transverse momentum of the charged lepton, $\slashed{E}_T$ denotes the missing transverse energy, and $\phi_{\ell\nu}$ represents the azimuthal angular separation between the charged lepton and the missing transverse momentum in the transverse plane. The transverse mass stands out as a pivotal parameter for disentangling signal and background events.

Both the ATLAS \cite{ATLAS:2019lsy, ATLAS:2017jbq} and CMS \cite{CMS:2018hff} collaborations utilize the transverse mass in their meticulous analyses of heavy charged boson production through mono-lepton channels to explore the frontiers of potential new physics. At the parton level, the charged lepton and the missing transverse momentum are consistently anti-aligned in the transverse plane. As a direct consequence, the transverse mass simplifies to precisely twice the transverse momentum of the charged lepton, i.e., $m_T = 2p_{\ell, T}$. This exact mathematical relationship assumes a preeminent role in experimental analyses at particle colliders, as it empowers researchers to distinguish effectively between signal and background processes.

In Fig.~\ref{fig:LX:MT:LHC13}(a), the production rate undergoes a precipitous decline beyond the resonant peak of the transverse mass $m_T$. In contrast, signal distributions exhibit an elongated tail at higher $m_T$ values. This stark disparity in the distribution profiles unequivocally demonstrates that the transverse mass $m_T$ emerges as an outstanding observable for differentiating signal events from background noise in the realm of particle physics experiments.

To validate the simulated distribution of the transverse mass, we conduct a comparison with the ATLAS results, which are derived from a substantially larger luminosity of 139 \ifb. Our event - selection protocol mandates that both the $\slashed{E}_T$ and the $p_{e, T}$ surpass $65\gev$. Furthermore, the charged lepton's rapidity is restricted to the range $|\eta_e| < 2.5$, with the barrel - endcap transition regions where $1.37 < |\eta_e| < 1.52$ being excluded and transverse mass  $m_T > 130\gev$.
The results of this validation are presented in Fig.~\ref{fig:LX:MT:LHC13}(b). In our analysis, the outcomes are normalized via a universal scale factor, $\epsilon_D = 0.65$, to bring them in line with the total count of irreducible background events documented in the experimental data. It's obviously that an excellent match in the transverse mass distribution and it is clear that within the experimental
uncertainty our simulation is valid. This remarkable congruence validates that the normalization approach not only succeeds in matching the total event counts but also in accurately portraying the differential distributions.

Fig.~\ref{fig:valid:mj:LHC13} to Fig.~\ref{fig:valid:mj:LHC:pers} illustrates the projected exclusion limits for LHC13, LHC14 and LHC25 at a 95\% confidence level (C.L.) within the $m_\chi-\varLambda$ parameter space respectively. These exclusion limits relevant to the LHC experaments are ascertained via the following $\chi^2$ computation:
\bee
\chi^2  =  \sum_{i}
\left[  \frac{ \epsilon_D \cdot N^{\rm S}_{i} }{ \sigma^{\rm ATLAS}_{i} } \right]^2\,,
\ene
In this context, $\sigma^{\rm ATLAS}_{i}$ stands for the experimental uncertainty associated with the $i$-th bin, as reported by the ATLAS research team, and $N^{\rm S}_{i}$ denotes the amount of signal events in that bin.
The factor $\epsilon_D = 0.65$, as previously expounded upon, is utilized to compensate for detector efficiency. This analytical strategy facilitates a stringent comparison and validation against the foreseen experimental outcomes at the LHC.
\begin{figure}[h]
\centering
\includegraphics[width=0.48\textwidth]{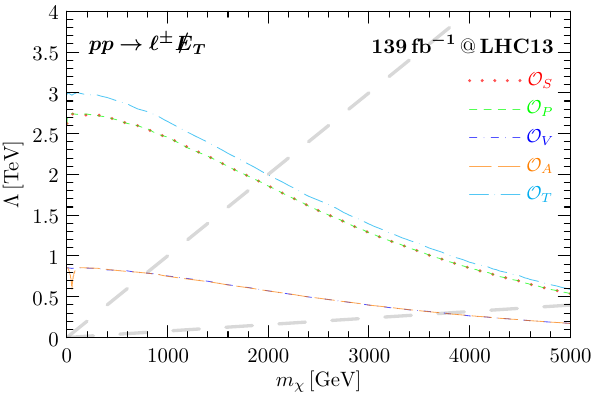}
\caption{\it
The expected exclusion limits at a 95\% confidence level (C.L.) are presented for the Large Hadron Collider (LHC) operating at a center-of-mass energy of $\sqrt{s}=13\tev$ and with a total integrated luminosity of $\mathcal{L}=139\fb^{-1}$.
}
\label{fig:valid:mj:LHC13}
\end{figure}
The mono-lepton process demonstrates a pronounced sensitivity to tensor operators. For a massless dark fermion, it can probe up to approximately $3\tev$ for LHC13. This elevated sensitivity stems from a larger number of events triggered by tensor operators in the high-$m_T$ region, as shown in Fig.~\ref{fig:LX:MT:LHC13}.
Conversely, the constraints on (axial-)vector operators are relatively weaker. When $m_\chi \approx 0$, the lower detection limit is around $1\tev$. Nevertheless, these differences become less significant as
the mass of the dark fermion rises. It is foreseeable that with higher luminosity and collision energy, the greater exclusion limits can be displayed, which is shown in right panel of Fig.~\ref{fig:valid:mj:LHC:pers}. Given that constraints on the energy scales depend on the Wilson coefficients of possible new physics at the electroweak scale, two thick gray-dashed lines with  \(\varLambda_i = m_\chi\)  and  \(\varLambda_i = m_\chi/4\pi\)  have been added for reference.

For the mono-lepton production process driven by four-fermion contact operators,
the signal cross-sections increase as the CoM energy rises.
In stark contrast, the corresponding background cross-sections decrease
as the CoM energy goes up.
Given the remarkable improvements in CoM energy and luminosity,
the future renditions of the LHC \cite{CidVidal:2018eel}
and the Super Proton-Proton Collider (SppC) \cite{CEPCStudyGroup:2023quu,CEPCStudyGroup:2018ghi} are excellently positioned to present substantial benefits in the exploration of DM via these operators.
The corresponding kinematic cuts for the three production processes are meticulously presented in Tab.~\ref{tab:hptc}.
\begin{table}[th]
\renewcommand\arraystretch{1.44}
\begin{center}\small
\begin{tabular}{ c | c  c  c  }
\hline\hline
{\rm Process}   & ~~~14~TeV, 3\iab ~~~ & ~~~~~25~TeV, 20\iab ~~~~~ &
\\\hline
\multirow{3}{*}{ $pp\to e^\pm \met$ }
&
\multicolumn{3}{c}{$\big|\eta_{e}\big| \leqslant 2.5$}
\\
& $p_{T,e} \geqslant 65\gev$ &  $p_{T,e} \geqslant 150\gev$ %&  $p_{T,e} \geqslant 450\gev$
\\
& $ \met \geqslant 65\gev$  &  $ \met \geqslant 150\,\gev$  &  %$ \met \geqslant 450\,\gev$
\\\hline\hline
\end{tabular}
\caption{\it The configurations of the LHC upgrades and the corresponding kinematic cuts at the parton level are presented.}
\label{tab:hptc}
\end{center}
\end{table}
\begin{figure}[h]
\centering
\includegraphics[width=0.48\textwidth]{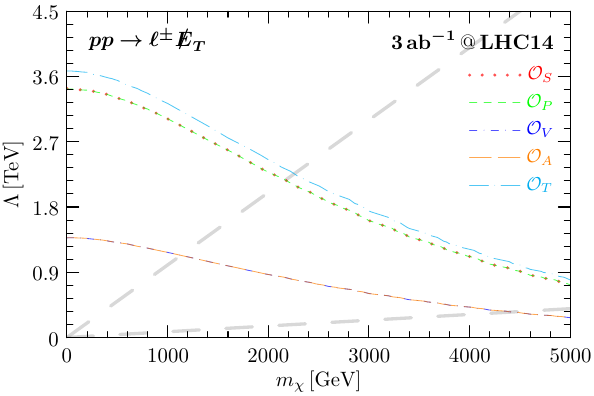}
\,
\includegraphics[width=0.48\textwidth]{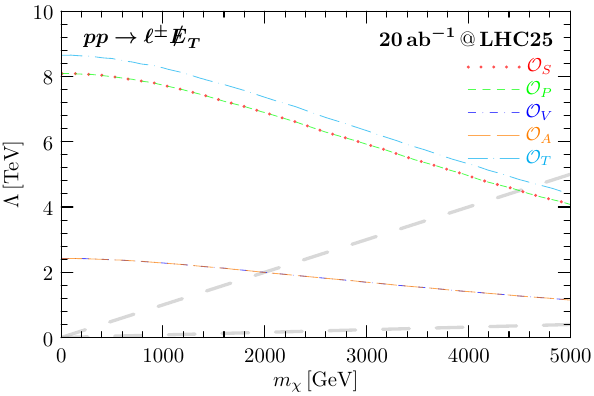}
\caption{\it
The expected exclusion limits at a 95\% confidence level are shown for the LHC
when it operates at a CoM energy of $\sqrt{s} = 14\tev$ and $25\tev$.
}
\label{fig:valid:mj:LHC:pers}
\end{figure}

\section{Absorptions at Nuclear Target}
\label{sec:Absor}
Studying non - trivial nucleon - level interactions induced by parton - level operators is a promising approach for direct DM detection. We focus on the inelastic scattering process where a dark fermion is completely converted to a neutrino, and its mass is fully absorbed to form the final nucleus \cite{Dror:2019onn,Dror:2019dib}:
\bee
\label{eq:abs:proc:nuc}
\chi(p_\chi) + A(p_{in}) \rightarrow \nu(p_\nu) + A(p_{fi})
\ene
Here, $p_{in}$ and $p_{fi}$ denote the momenta of the initial and final states, and the mass number $A$ represents the input and output nuclear targets.
The differential scattering rate per nuclear recoil energy $E_R$ is given by
\bee
\label{eq:dd:drt}
\frac{\mathrm{d} R_A}{\mathrm{d} E_R}=N_A n_\chi\int \mathrm{d}^3 v f_{\mathrm{E}}(\boldsymbol{v}_\chi, t) v_\chi\frac{\mathrm{d} \sigma_A}{\mathrm{d} E_{\mathrm{R}}}
\ene
where $N_A$ is the number of nuclear targets. Usually, the total rate is the sum of individual contributions from each isotope. The local number density of DM is $n_\chi=\rho_\chi/m_\chi$, with a local energy density $\rho_\chi \simeq 0.3 \,{\rm GeV/cm}^3$. Moreover, $\mathrm{d}\sigma_{\rm A}/\mathrm{d} E_{\rm R}$ is the differential cross - section of the absorption process in \eqref{eq:abs:proc:nuc}, and $f_{\mathrm{E}}(\boldsymbol{v}_\chi, t)$ is the velocity distribution of the absorbed DM.

In the absorption process, the leading-order recoil energy of the nucleus is approximately proportional to the square of the DM mass $m_\chi^2$, where $m_\chi$ is substantially smaller than the nucleus mass $m_{A}$, and this relationship is given by
$E_{R}^0 = m_\chi^2/2(m_{A}+m_\chi) \approx m_\chi^2/2m_{A}$.
As a result, the differential cross - section peaks sharply at $E_{R}^0$,
which can be modeled by the Dirac Delta function:
\bee
\label{eq:dd:dxst}
\frac{\mathrm{d}\sigma_{A}}{\mathrm{d} E_{\rm R}}=\frac{ \overline{\big| \mathcal{M}_{A} \big|^2} }{16 \pi m_{A}^2 v_\chi  }\delta\big( E_R - E_R^0 \big)
\ene
Here, $\mathcal{M}_{A}$ is the helicity amplitude of the process in \eqref{eq:abs:proc:nuc}, defined as
\bee
\overline{|\mathcal{M}_A|^2}\equiv\frac{1}{2 s_{\chi}+1} \frac{1}{2 J+1}\sum_{\substack{\text { initial \& } \\ \text { final spins }}}|\mathcal{M}_A|^2
\ene
where $J$ is the total spin of the nucleus and $s_{\chi} = 1/2$ is the spin of the dark fermion.
For non-relativistic DM, the dominant contribution to the scattering amplitude $\mathcal{M}$ occurs when the DM momentum $p_\chi \to 0$. This indicates that the leading-order amplitude of the absorption process is independent of the DM velocity. Thus, the velocity term in \eqref{eq:dd:drt} can be treated separately, simplifying the theoretical analysis of DM interactions.
The DM velocity distribution function $f_{\mathrm{E}}(\boldsymbol{v}_\chi, t)$ is normalized to 1, and the differential scattering rate can be rewritten as
\bee
\frac{\mathrm{d} R_A}{\mathrm{d} E_{\mathrm{R}}}=\frac{ N_A  n_\chi }{16 \pi m_{A}^2  }\overline{\big| \mathcal{M}_A \big|^2}\delta\big( E_R - E_R^0 \big)  \varTheta(E_{\rm R} - E_{\rm R}^{\rm th})
\ene
where $E_{\rm R}^{\rm th}$ is the recoil energy threshold, and the step function $\varTheta(E_{\rm R} - E_{\rm R}^{\rm th})$ accounts for the minimum observable energy in an experiment.

For simplicity, this analysis focuses solely on the coherent process. Effective operators with different Lorentz structures can induce both spin-independent (SI) and spin-dependent (SD) interactions, which are differentiated by a coherence factor $C_N$ at the nuclear level.
The corresponding differential rate of recoil events can be written as
\bee
\label{eq:nucleus:ff}
\mathcal{M}_A(q^2)=\sum_{N = p, n} F_{\rm Res}^{N} (q^2)\mathcal{M}_{\rm PLN}^{N}(q^2)=\sum_{N = p, n} F_{\rm Res}^{N} (q^2) C_N \mathcal{M}_{N}(q^2)
\ene
where the squared momentum transfer $q^2$ is defined as $(p_{fi}-p_{in})^2=(p_\nu - p_\chi)^2$. The amplitude $\mathcal{M}_{\rm PLN}^{N}(q^2)$ describes the DM scattering off a point - like nucleus (PLN) and is derived from the amplitude for scattering off a single nucleon, $\mathcal{M}_{N}(q^2)$, normalized to the nuclear mass $m_A$. Moreover, $F_{\rm Res}^{N}(q^2)$ represents the corresponding nuclear response functions that characterize the nuclear effects on the scattering process at different momentum transfers.
The coherence of nucleon spin states differs between SI and SD interactions, influencing the value of the coherence factor $C_N$. In SI processes, $C_N$ equals $Z$ for protons and $A - Z$ for neutrons. In contrast, for SD interactions, the coherence factor simplifies to $C_N = 1$, indicating a uniform contribution from both protons and neutrons without interference.

\subsection{Spin-Independent Absorption}
\label{sec:DD:SI}
As illustrated in \eqref{eq:nucleus:ff}, the SI absorption process at the nuclear scale is directly associated with the nucleon - level amplitude:
\bee
\mathcal{M}_A = \sum_{N = p, n} F_{SI}^{N}(q^2)  C_N \mathcal{M}_{N}
\ene
The SI-specific response function $F_{SI}^{N}(q^2)$ can typically be characterized by the Helm form factor $F_{\rm Helm}(q^2)$ of the target nucleus. We follow the formulation detailed in Refs. \cite{Helm:1956zz,Lewin:1995rx}.
Furthermore, when accounting for the conservation of isospin symmetry, which indicates that the disparity between the scattering amplitudes off a proton and a neutron is insignificant, we find that $\mathcal{M}_p = \mathcal{M}_n$.
\begin{figure}[!h]
\centering
\includegraphics[width = 0.48\textwidth]{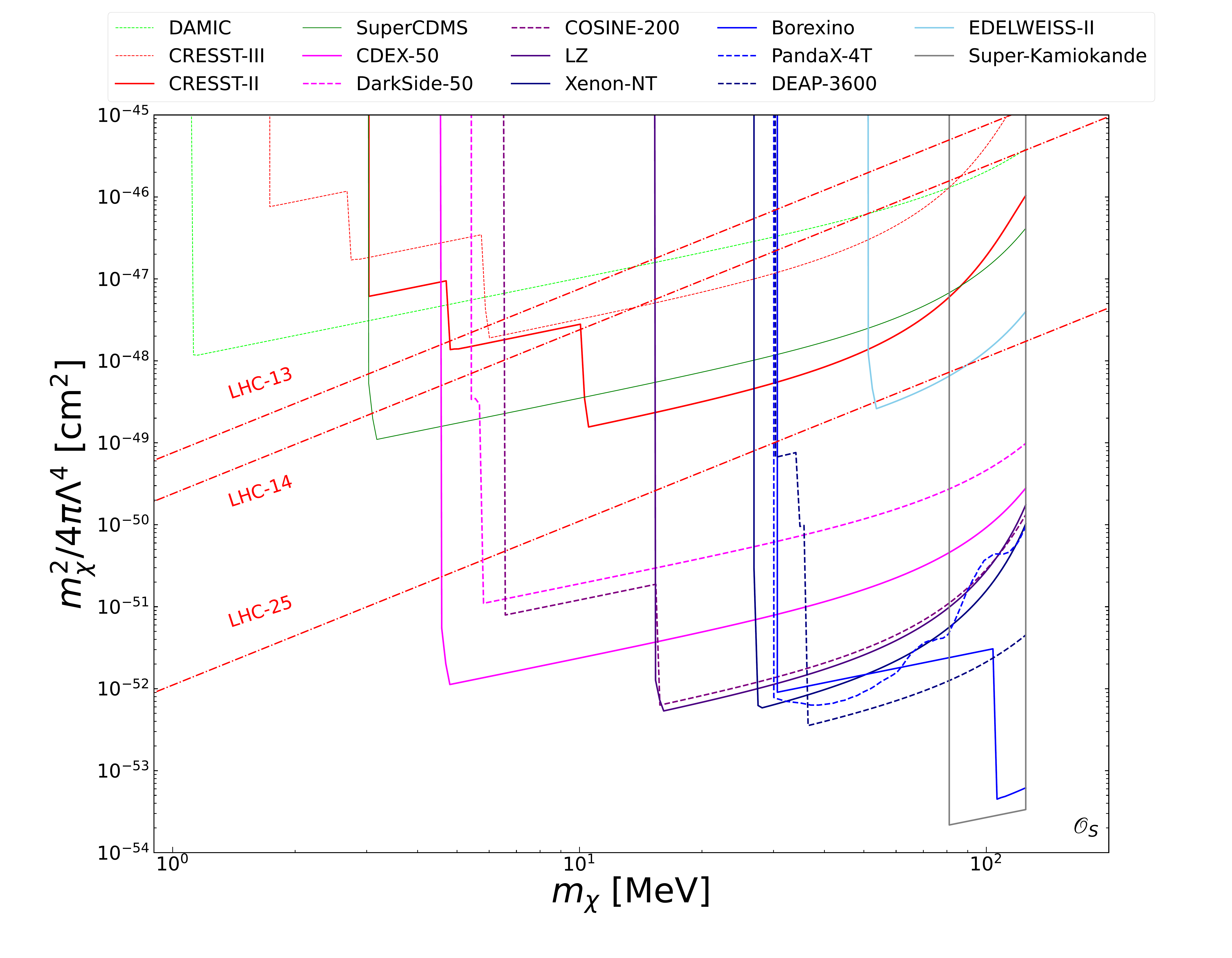}
\includegraphics[width = 0.48\textwidth]{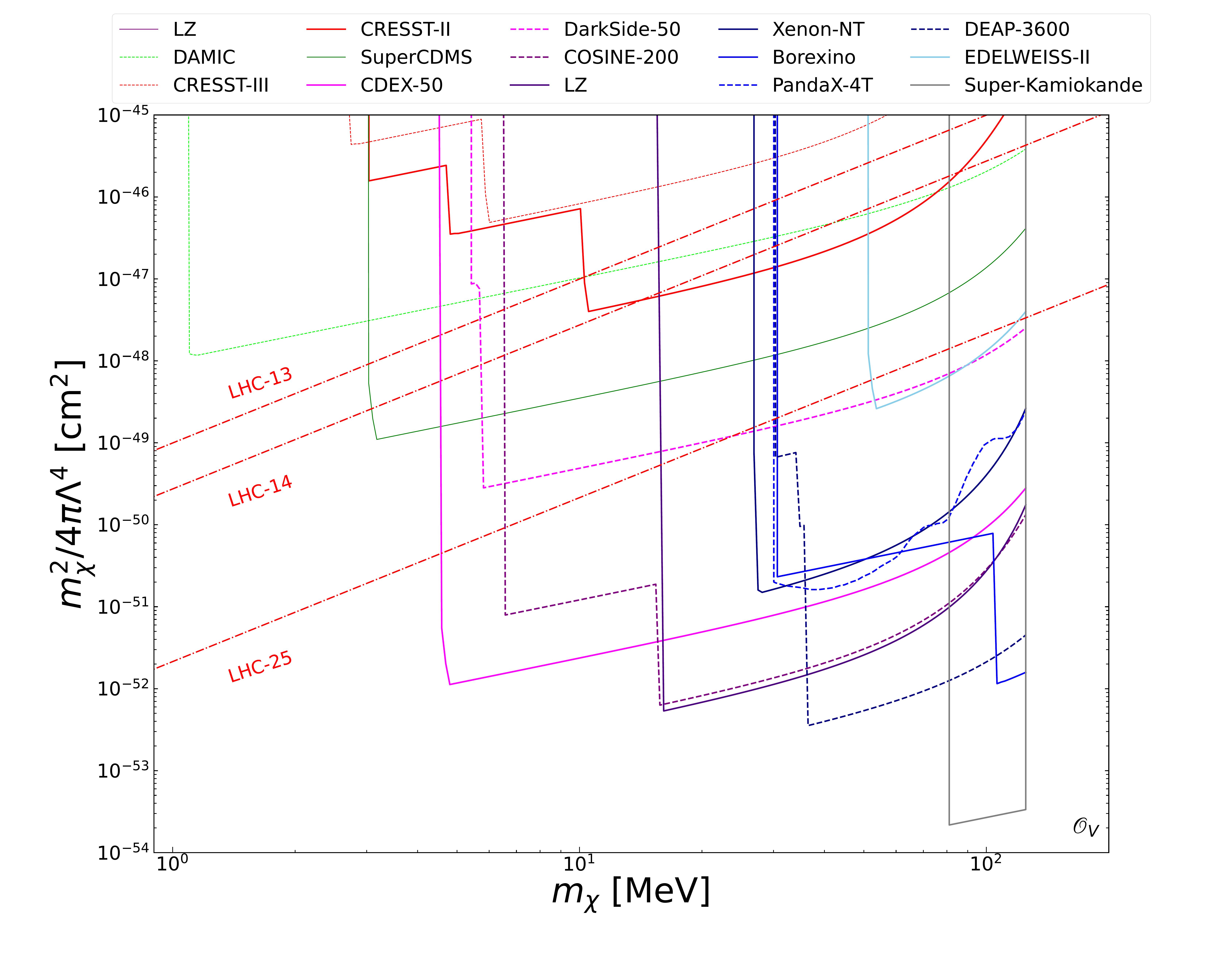}
\caption{\it
Excluded regions in the $m_\chi^2/4\pi\varLambda^4$ - $m_\chi$ plane for the scalar (\textbf{left panel}) and vector (\textbf{right panel}) operators.
}
\label{fig:SI:S}
\end{figure}

Based on the non-relativistic expansions in Refs. \cite{Ma:2024tkt}, the scalar and vector components defined in \eqref{eq:effo} can make substantial contributions to SI absorption; thus, $\varGamma = S$ and $V$. Through simple calculations,
we obtain an excellent expression for the squared average of the amplitude:
\bea
\overline{|\mathcal{M}_{N}^{S}|^2} &=& \frac{4 m_{A}^2 m_\chi^2 }{ \varLambda^4_S } F_{S}^{N} \,, \quad\quad F_{S}^{N} = \sum_{q} \frac{ F^{q/N}_{S} }{m_q} \,,\\[3mm]
\overline{|\mathcal{M}_{N}^{V}|^2} &=& \frac{4 m_{A}^2 m_\chi^2 }{ \varLambda^4_V } F_{V}^{N}\,, \quad\quad F_{V}^{N} = \sum_{q} F^{q/N}_{1} \,,
\ena
Assuming that the DM couples universally to quarks,
we gather all relevant terms to determine the scattering rate as follows:
\bee
R_A^{\varGamma} = N_A  n_\chi  \frac{ m_{\chi}^2 }{4 \pi  \varLambda_{\varGamma} } \big[ A F_{\varGamma}^{N}(m_\chi^2) F_{\rm Helm}(m_\chi^2) \big]^2 \varTheta(E_{\rm R}^0 - E_{\rm R}^{\rm th}) \,,
\ene
where we use the approximation $q^2 \approx m_\chi^2$.
The excluded regions obtained in the $m_\chi^2/4\pi\varLambda^4$-$m_\chi$ plane for $\mathcal{O}_{S}$ and $\mathcal{O}_{V}$ are presented in Fig.~\ref{fig:SI:S}.
Evidently, the Super-Kamiokande collaboration imposes the strongest constraint on the SI scattering cross section for fermionic DM due to its large exposure. Moreover, compared with the LHC bounds, the absorption process always provides stronger constraints in the region $m_\chi \in [\sim10, \sim100]\mev$. We will explore the other experiments listed in Tab.~ \ref{tab:SD:Exposure}.

However, the high recoil energy threshold in Super-Kamiokande implies that it will only enforce strong constraints for heavy DM. Given that lighter nuclei are highly efficient in probing light DM candidates, we can project the scattering outcomes off individual hydrogen nuclei in future experiments within the mass range $m_\chi \in [\sim 0,\sim 1]\mev$. This projection assumes an energy threshold $E_{th}$ of 1 eV and an exposure of 100 kg.yr \cite{Dror:2019dib}, as depicted in Fig.~\ref{fig:SI:S:light} for $\mathcal{O}_S$ and $\mathcal{O}_T$.
\begin{figure}[!h]
\centering
\includegraphics[width = 0.6\textwidth]{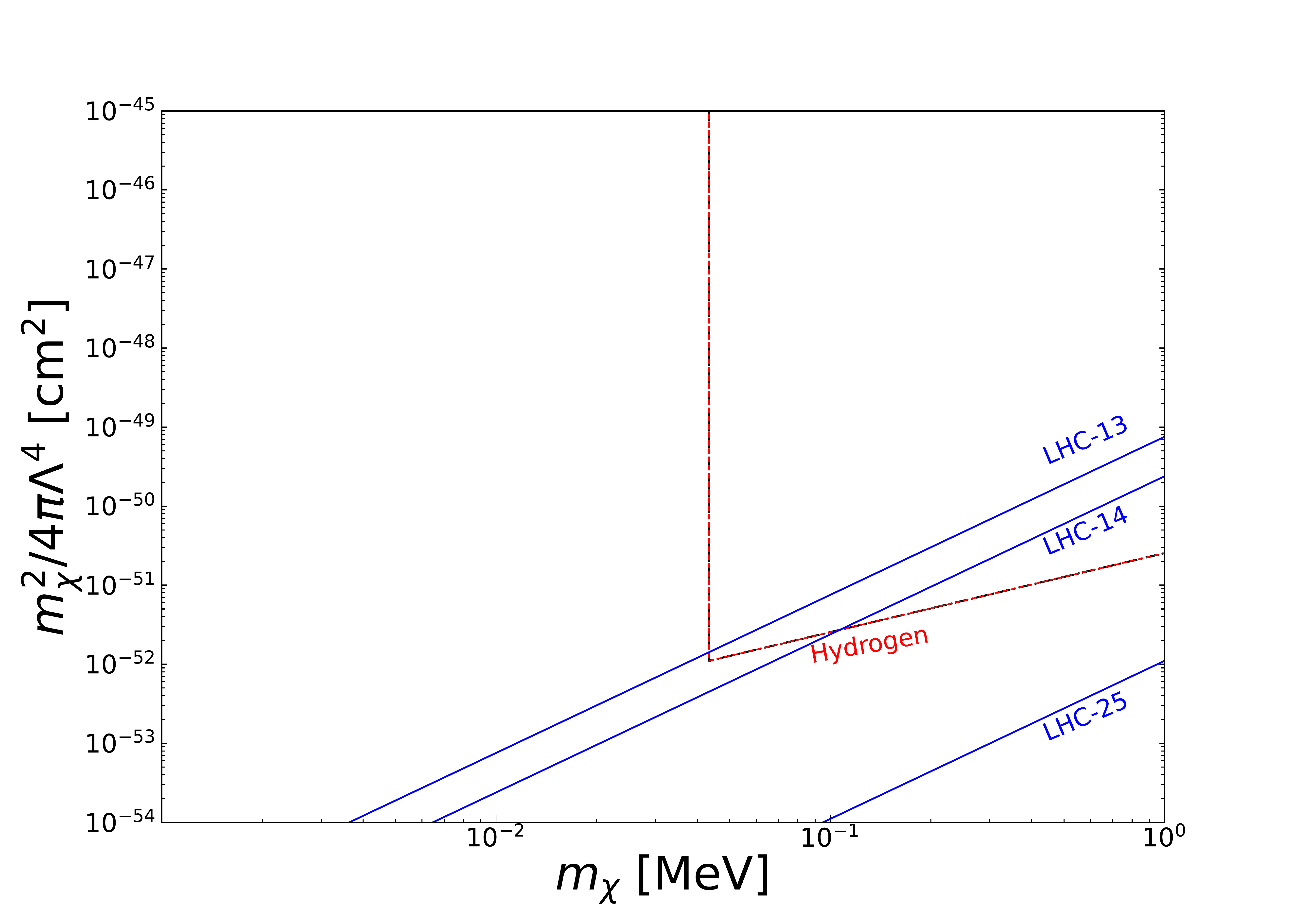}
\caption{\it
Excluded regions in the $m_\chi^2/4\pi\varLambda^4$ - $m_\chi$ plane for scalar and vector operators when considering hydrogen as the target nuclei.
}
\label{fig:SI:S:light}
\end{figure}
Evidently, the LHC imposes the most stringent direct constraints, highlighting the collider's effectiveness as a potent tool for detecting light DM.

\subsection{Spin-Dependent Absorption}
\label{sec:DD:SD}
The SD interactions can be triggered by pseudo-scalar and axial-vector operators. Generally, the leading order contribution of the pseudo-scalar operator is suppressed by an additional factor of $q/m_A \sim m_\chi/m_A$. The leading order nonrelativistic nucleon matrix elements for the pseudo-scalar and axial-vector operators can be written as follows:
\bea
\mathcal{O}^{N}_{P} &=& 2\sqrt{2} h_{\nu} m_\chi F_{P}^{N} \big[(\xi_{h_\nu}^{\nu})^\dag \xi^{\chi}_{h_\chi} \big] \big[(\omega_{h_{N'}}^{N'})^\dag  \big( \boldsymbol{q} \cdot \boldsymbol{s}\, \big) \,\omega_{h_N}^N \big] \,,\\
\mathcal{O}^{N}_{A} &=& 8\sqrt{2}  m_\chi m_A F_{A}^{N} \big[(\xi_{h_\nu}^{\nu})^\dag \boldsymbol{s}\, \xi^{\chi}_{h_\chi} \big] \big[(\omega_{h_{N'}}^{N'})^\dag  \boldsymbol{s} \,\omega_{h_N}^N \big] \,,
\ena
Given that the amplitude can be represented as the coherent sum of individual nucleon contributions, the corresponding total amplitude is as follows:
\bea\label{SD:Am}
\mathcal{M}_{A}^A &=& \sum_{N = p, n} \boldsymbol{s}_\chi \cdot \left\langle J, M'\left|\boldsymbol{S}_N\right| J, M\right\rangle \,,\\[1mm]
\mathcal{M}_{P}^A &=& \sum_{N = p, n} \mathbb{I}_\chi \; \big[ \left\langle J, M^{\prime}\left|\boldsymbol{S}_N\right| J, M\right\rangle \cdot \boldsymbol{q} \big] \,,
\ena
Here, $\boldsymbol{s}_\chi$ represents the spin operators of the DM or neutrino, and $\mathbb{I}_\chi$ is the identity operator of this spinor space. For a nuclear state with total spin $J$ and helicity $M$, $\boldsymbol{S}_N$ denotes the total nucleon spin operator, where $\boldsymbol{S}_p \equiv \sum_{i=\text{protons}} \boldsymbol{s}_{p_i}$ and $\boldsymbol{S}_n \equiv \sum_{i=\text{neutrons}} \boldsymbol{s}_{n_i}$ for protons and neutrons, respectively.
In a spinless nuclear state, where the total spin $J=0$, the above matrix elements clearly become zero. Tab.~\ref{tab:SD:Exposure} lists some typical experiments involving isotopes with non-zero total spin. These experiments effectively impose excellent constraints on the SD scattering cross section for fermionic DM in this study.

Upon summing over the final state spins and averaging over the initial states, the squared amplitudes are provided as follows:
\bea
\overline{\big| \mathcal{M}_{A}^A \big|^2} &=& \frac{1}{2} \sum_{N, N'} F_{4,4}^{NN'}= \sum_{N, N'}  \frac{1}{32} \Big[ F_{\Sigma'}^{NN'}(q^2) + F_{\Sigma''}^{NN'}(q^2)  \Big] \,,\\
\label{eq:ls:sm}
\overline{\big| \mathcal{M}_{P}^A \big|^2 } &=& \sum_{N, N'} F_{10,10}^{NN'}= \sum_{N, N'}  \frac{q^2}{4} F_{\Sigma''}^{NN'}(q^2)  \,,
\ena
Here, the additional factor of $1/2$ is attributed to the fact that only left-handed neutrinos partake in the scattering.

The explicit expressions of $F_{\Sigma'}^{NN'}(q^2)$ and $ F_{\Sigma''}^{NN'}(q^2)$ are defined in Ref.~\cite{Fitzpatrick:2012ix}. The approximation of zero momentum transfer can be concisely written as:
 \bee
\label{eq:ssr}
 \sum_{M, M'}
\left\langle J, M\left|S_{N', i}\right| J, M'\right\rangle
\left\langle J, M'\left|S_{N, j}\right| J, M\right\rangle
=
\frac{(J + 1)(2 J + 1)}{3 J}
\mathbb{S}_N \mathbb{S}_{N'} \delta_{i j}
\ene
Here, $\mathbb{S}_N \equiv \left\langle J, J \left|S_N^z\right| J, J\right\rangle$ denotes the expectation values of the nuclear spin operator for states with maximal angular momentum. The response functions involving $\mathbb{S}_N$ are then given as:
\bea
\label{eq:ss:boundary}
F_{4,4}^{NN'} &=& \frac{J + 1}{4J} \mathbb{S}_N \mathbb{S}_{N'} \,, \quad F_{\Sigma'}^{NN'} + F_{\Sigma''}^{NN'} = \frac{4(J + 1)}{J}  \mathbb{S}_N \mathbb{S}_{N'} \,,\\[3mm]
F_{10,10}^{NN'} &=& \frac{J + 1}{3J} q^2 \mathbb{S}_N \mathbb{S}_{N'} \,, \quad F_{\Sigma''}^{NN'} = \frac{4(J + 1)}{3J}  \mathbb{S}_N \mathbb{S}_{N'} \,.
\ena
The $q^2$ dependence of the response function $F_{10,10}^{NN'}$ can be factored out using the approximation $F_{10,10}^{NN'}(q^2 = 0) = \frac{q^2}{4} F_{\Sigma''}^{NN'}(q^2 = 0)$. Moreover, the model - dependent parameter $\mathbb{S}_N$ is determined via detailed calculations within realistic nuclear models. We list the values of $\mathbb{S}_N$ in Tab.~ \ref{tab:exps}. For more details on $\mathbb{S}_N$, we direct the reader to Ref.~\cite{Stadnik:2014xja}.

In our study, when calculating the scattering rate, we assume that either a proton or a neutron takes part in the interaction. Additionally, only the nucleon with the largest spin expectation value contributes to the scattering rate. After accounting for the phase space factor, the scattering rate is then given by:
\bea
\label{eq:SD:rate:p}
R_A^{P} &=& \frac{ N_A  n_\chi  m_{\chi}^4 }{8 \pi  \varLambda_{P}^4 m_{A}^2 } \varTheta(E_{\rm R}^0 - E_{\rm R}^{\rm th})\sum_{N,N' = p, n} F_{P}^{N} F_{P}^{N'} \big[ F_{\Sigma''}^{NN'} \big] \,,\\[3mm]
R_A^{A} &=& \frac{ N_A  n_\chi  m_{\chi}^2 }{4 \pi  \varLambda_{A}^4  } \varTheta(E_{\rm R}^0 - E_{\rm R}^{\rm th})\sum_{N,N' = p, n} F_{A}^{N} F_{A}^{N'} \big[ F_{\Sigma'}^{NN'} + F_{\Sigma''}^{NN'} \big] \,.
\ena

Figure \ref{fig:SD:P} shows the excluded regions from SD scatterings in the $m_\chi^2/4\pi\varLambda^4$-$m_\chi$ plane for the experiments listed in Tab.~\ref{tab:SD:Exposure}.
\begin{figure}[htpb!]
\centering
\includegraphics[width = 0.49\textwidth]{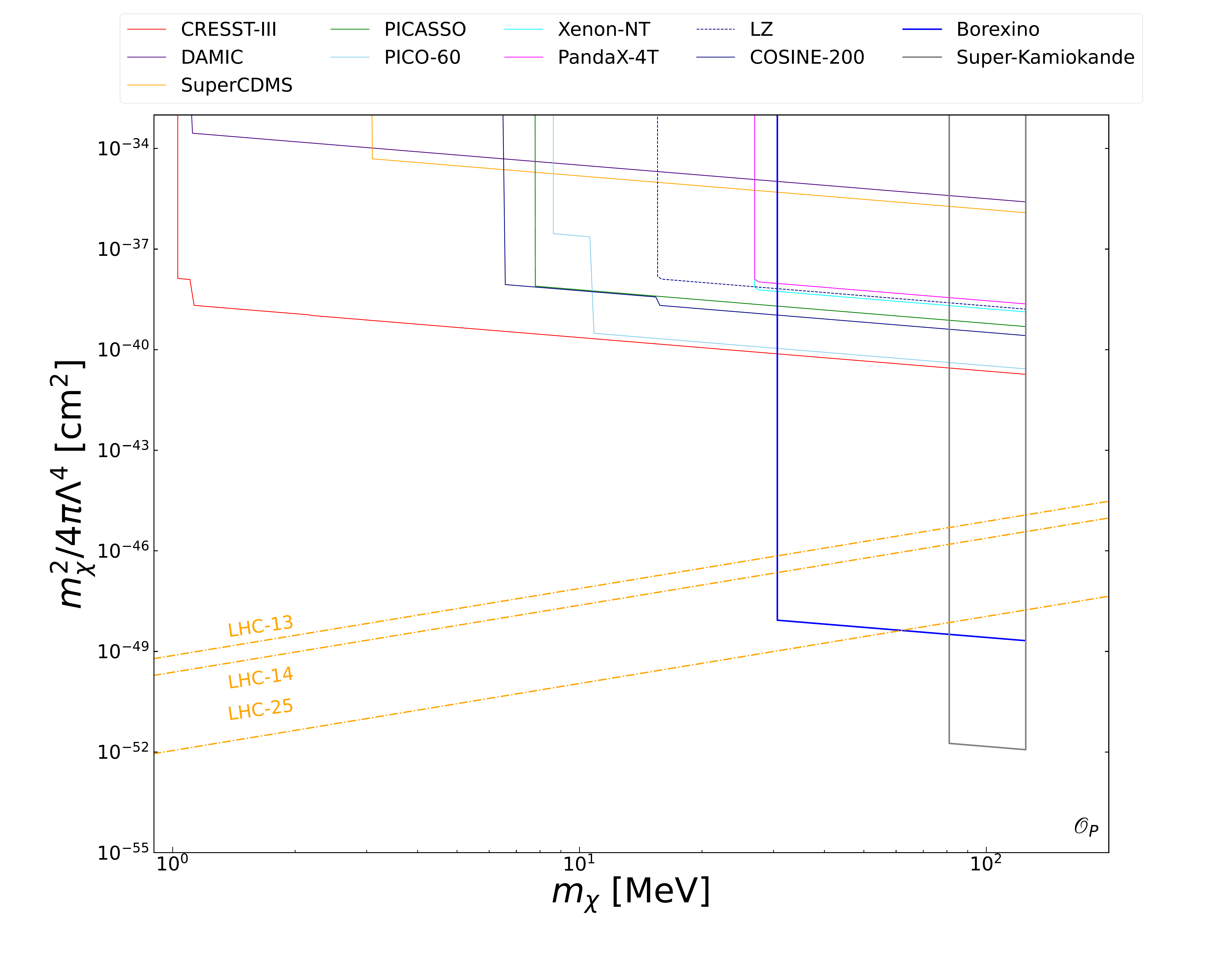}
\includegraphics[width = 0.49\textwidth]{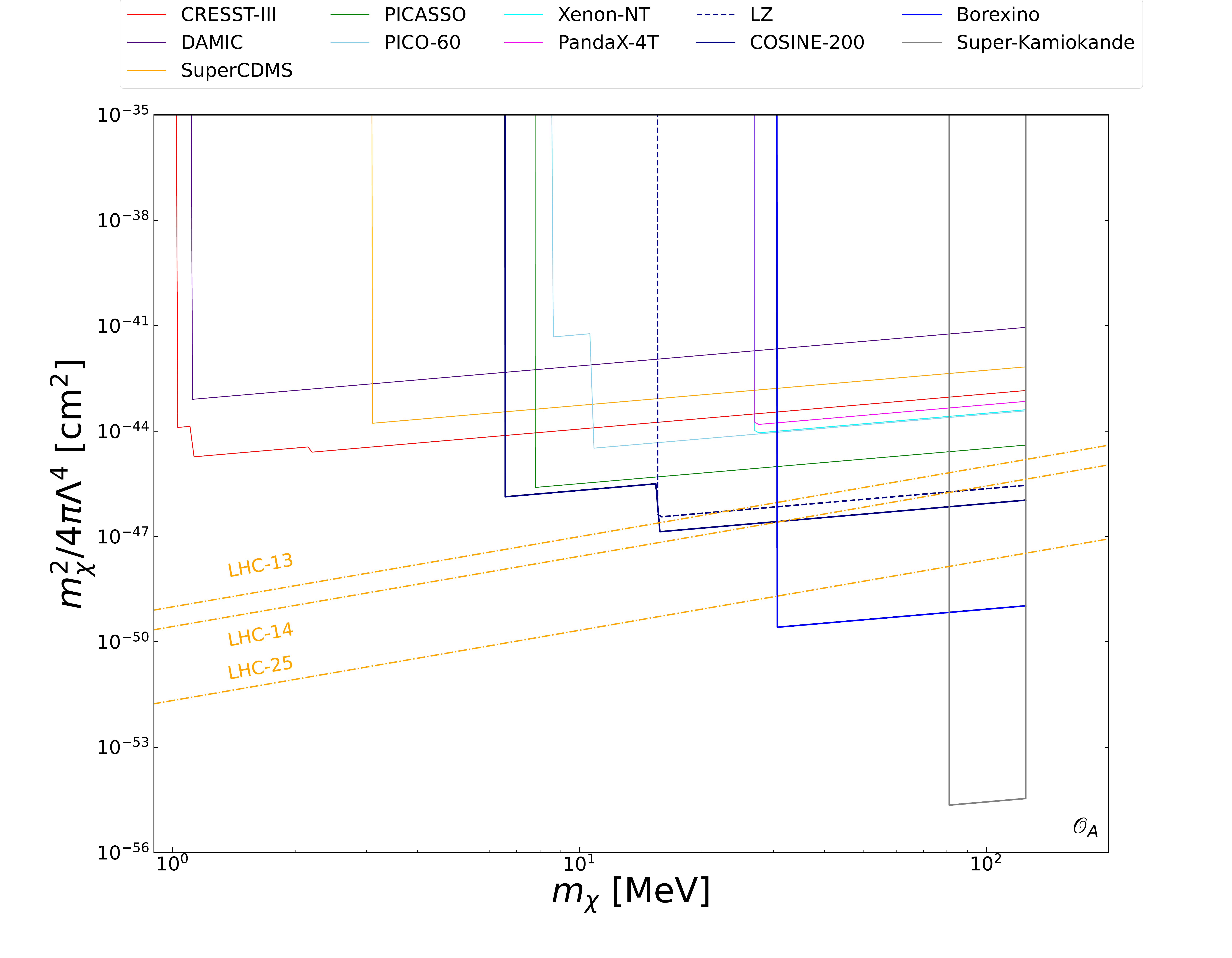}
\caption{\it
Excluded regions in the $m_\chi^2/4\pi\varLambda^4$ - $m_\chi$ plane from SD scatterings for the pseudo - scalar (left) and axial - vector (right) operators.}
\label{fig:SD:P}
\end{figure}
In nontrivial nucleon level interactions, the constraints from the pseudo-scalar operator $\mathcal{O}_P$ are far less stringent than those from the axial-vector operator $\mathcal{O}_A$. Notably, the Super-Kamiokande collaboration still imposes the tightest constraints compared to other direct detection experiments. The reasons can be summarized as follows:
Firstly,  \eqref{eq:SD:rate:p} shows that the scattering rate of $\mathcal{O}_P$ is suppressed by a factor of $m^2_\chi/m^2_A$. Thus, the scattering rate of $\mathcal{O}_A$ is significantly larger than that of $\mathcal{O}_P$.
Secondly, the constraints from SD experiments using heavier nuclei are much weaker than those using lighter nuclear targets. This difference is clearly shown in the results of the Borexino and Super-Kamiokande experiments. These experiments not only have larger exposures but also use lighter isotopes, as detailed in Tab.~\ref{tab:SD:Exposure}. These features greatly enhance the experiments' sensitivity to SD interactions. As a result, the constraints from the Borexino and Super-Kamiokande experiments are stronger than those of most experiments and the LHC bounds.

\begin{figure}[!h]
\centering
\includegraphics[width=0.6\textwidth]{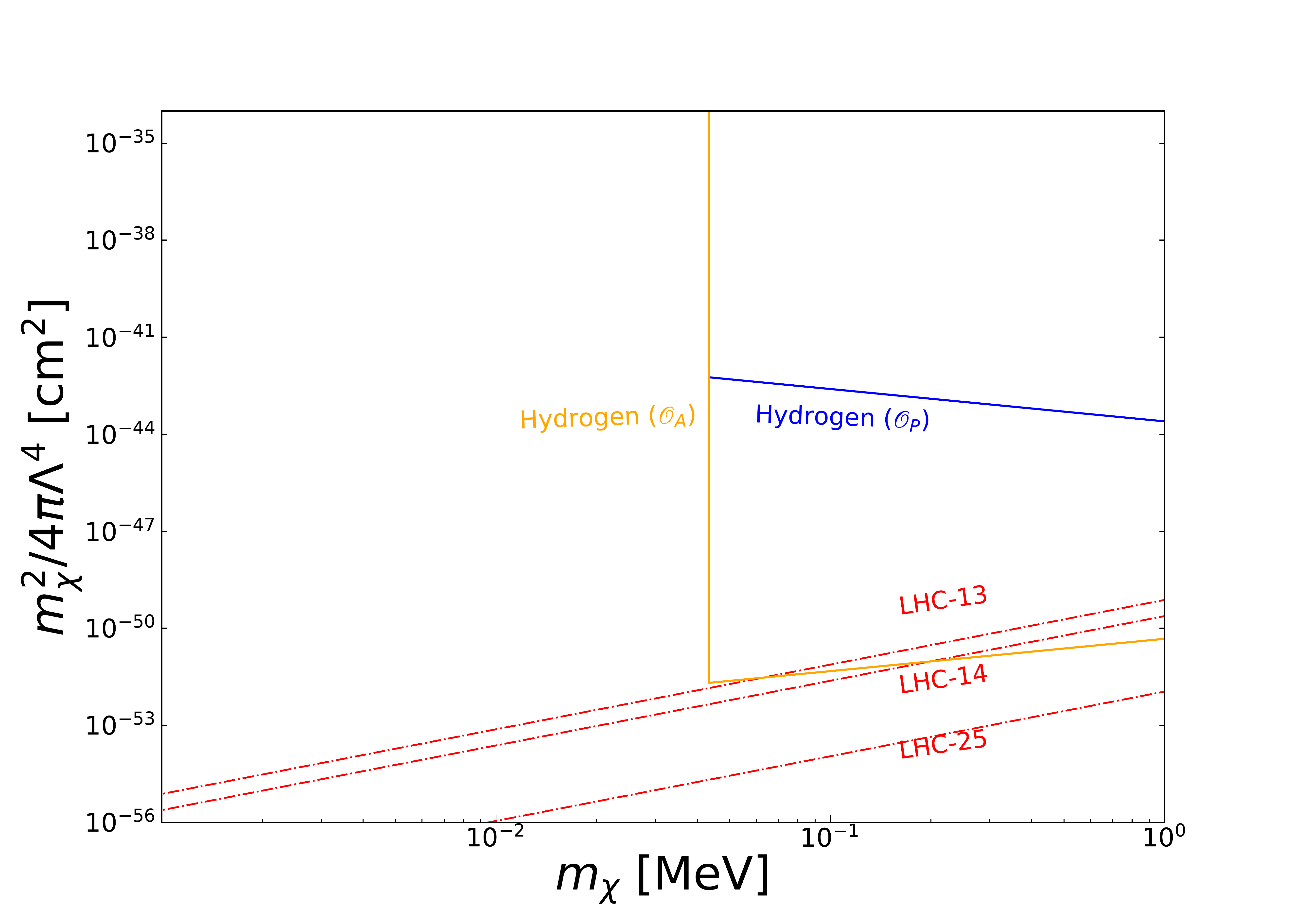}
\caption{\it
Excluded regions in the $m_\chi^2/4\pi\varLambda^4$-$m_\chi$ plane due to SD scatterings for the pseudo-scalar and axial-vector operators with hydrogen as the target.}
\label{fig:SD:P:light}
\end{figure}
We also provide constraints from hydrogen targets for the operators $\mathcal{O}_P$ and $\mathcal{O}_A$ in the mass range where $m_\chi \lesssim 1$ \mev. As shown in Fig.~\ref{fig:SD:P:light}, the LHC has a clear advantage in the low mass region when compared to direct detection experiments.

\subsection{Absorption for the Tensor Operator}
\label{sec:DD:Ten}
The quark level tensor operator $\mathcal{O}_T$ can induce both SI and SD interactions at the nucleon level. Therefore, the total interaction amplitude is the combined sum of these two contributions:
\bee
\mathcal{M}_{A,T} = \mathcal{M}_{A,T}^{SD} + \mathcal{M}_{A,T}^{SI} \,.
\ene
Similar to the SD absorption case in \eqref{SD:Am}, the scattering amplitude for the tensor operator can be expressed as:
\bee
\mathcal{M}_{A,T}^{SD} = 16\sqrt{2}  m_\chi m_A \sum_{N = p,n} F_{T,SD}^{N} \big[ \boldsymbol{s}_\chi \cdot \left\langle J, M'\left|\boldsymbol{S}_N\right| J, M\right\rangle \big]\,,
%\calo_{SS}
\quad
F_{T,SD}^{N} = \sum_{q} \frac{ F^{q/N}_{T,0} }{m_q} \,.
\ene
For the SI interaction, the corresponding scattering amplitude is:
\bee
\mathcal{M}_{A,T}^{SI} = 4 \sqrt{2} m_\chi  \sum_{N = p,n} F_{T,SI}^{N} C_N F_{SI}^{N} \big[ \boldsymbol{s}_\chi \cdot \boldsymbol{q} \big] \delta_{M'M}\,,
\quad
F_{T,SI}^{N} = \sum_{q} \frac{ F^{q/N}_{T,1} + 2F^{q/N}_{T,2} }{m_q} \,.
\ene
Therefore, the total average squared amplitude is given by:
\bee
\overline{\big| \mathcal{M}_{T, L} \big|^2} = \overline{\big| \mathcal{M}_{T, SD} \big|^2} + \overline{\big| \mathcal{M}_{T, SI} \big|^2} + \overline{\mathcal{M}_{T, SD}^\dag  \mathcal{M}_{T, SI} + \mathcal{M}_{T, SI}^\dag  \mathcal{M}_{T, SD}  } \,.
\ene
After summing over the helicity states of both DM particles and neutrinos, the interference term can be reformulated as follows:
\bee
\overline{\Re\big\{ \mathcal{M}_{T, SD}^\dag  \mathcal{M}_{T, SI} \big\} } \propto \sum_{M} \left\langle J, M \left| \boldsymbol{q} \cdot \boldsymbol{S}_N\right| J, M\right\rangle \,.
\ene

However, since the nuclear target is always unpolarized and the orientation of momentum transfer is isotropic, the interference term makes a negligible contribution to the total event rate. This is because the overall scattering rate represents an incoherent sum of the SD and SI contributions:
\bee
R_A^{T} = R_{A, SD}^{T} + R_{A, SI}^{T} \,,
\ene
where the detailed expressions for the SD and SI scattering rates are as follows:
\bea\label{AM:Ten}
R_{A, SD}^{T} &=& \frac{ 16 N_A  n_\chi  m_{\chi}^2 }{\pi  \varLambda_{T}^4  } \varTheta(E_{\rm R}^0 - E_{\rm R}^{\rm th}) \sum_{N,N' = p, n} F_{T,SD}^{N} F_{T,SD}^{N'} \big[ F_{\Sigma'}^{NN'} + F_{\Sigma''}^{NN'} \big]  \,,\\
R_{A, SI}^{T} &=& N_A  n_\chi  \frac{ m_{\chi}^4 }{4 \pi  \varLambda_{T} m_{A}^2 } \big[ A F_{T,SI}^{N} F_{\rm Helm}(m_\chi^2) \big]^2 \varTheta(E_{\rm R}^0 - E_{\rm R}^{\rm th}) \,.
\ena
From \eqref{AM:Ten}, it is evident that the SI contribution is suppressed by a factor of $m^2_\chi/m^2_A$, similar to the pseudo-scalar contribution to the SD scattering discussed in the previous section. The left panel of the Fig.~\ref{fig:SD:T} shows the excluded regions in the $m_\chi^2/4\pi\varLambda^4$ - $m_\chi$ plane due to SD scatterings for the tensor operator. Clearly, the Borexino and Super-Kamiokande experiments are still the most promising methods for detecting the absorption signal. Moreover, the constraints imposed by the LHC cover nearly all those derived from direct detection experiments and are particularly dominant in the low DM mass region. This superiority is clearly shown in the right panel of the Fig.~\ref{fig:SD:T}.
\begin{figure}[!h]
\centering
\includegraphics[width = 0.49\textwidth]{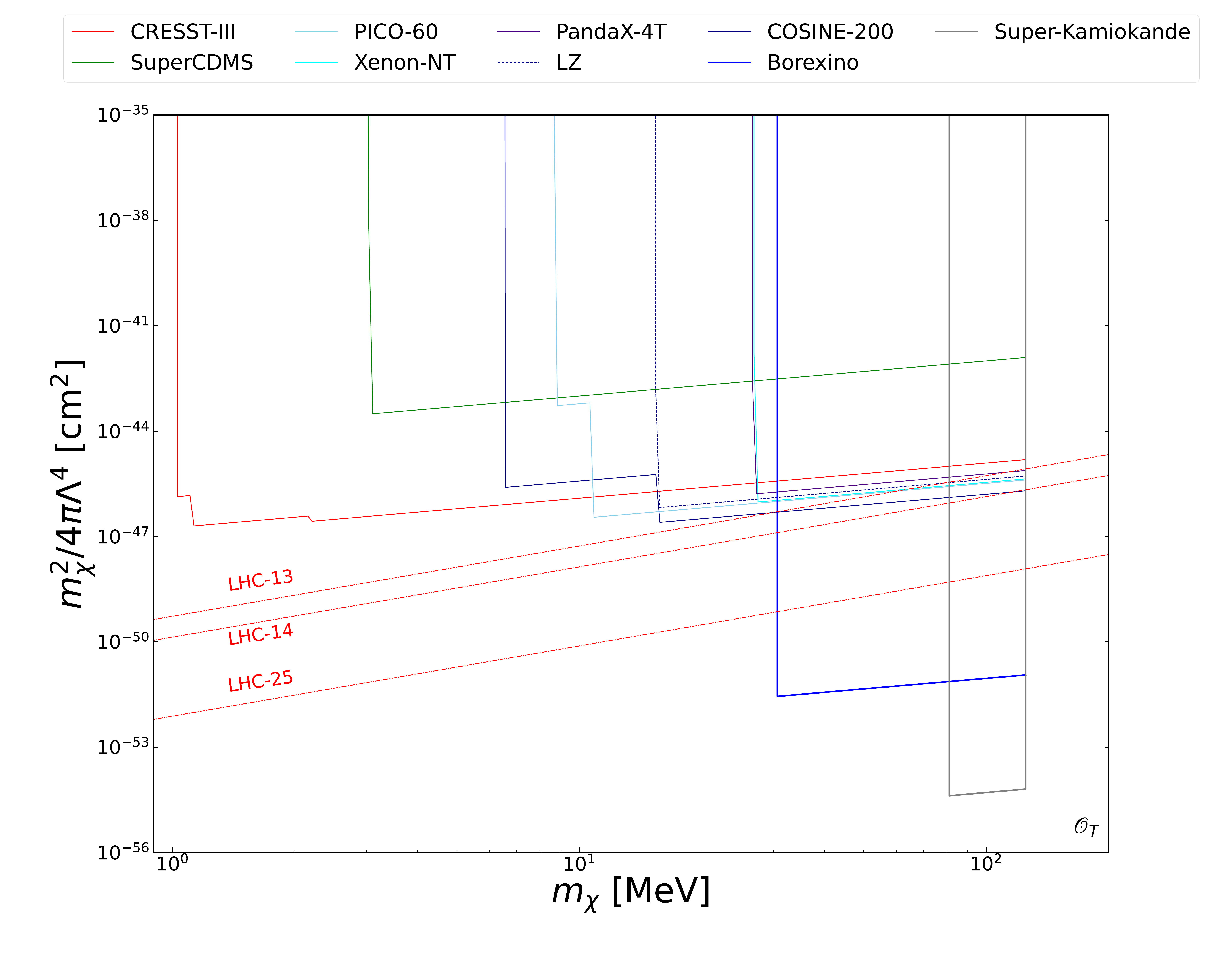}
\includegraphics[width = 0.49\textwidth]{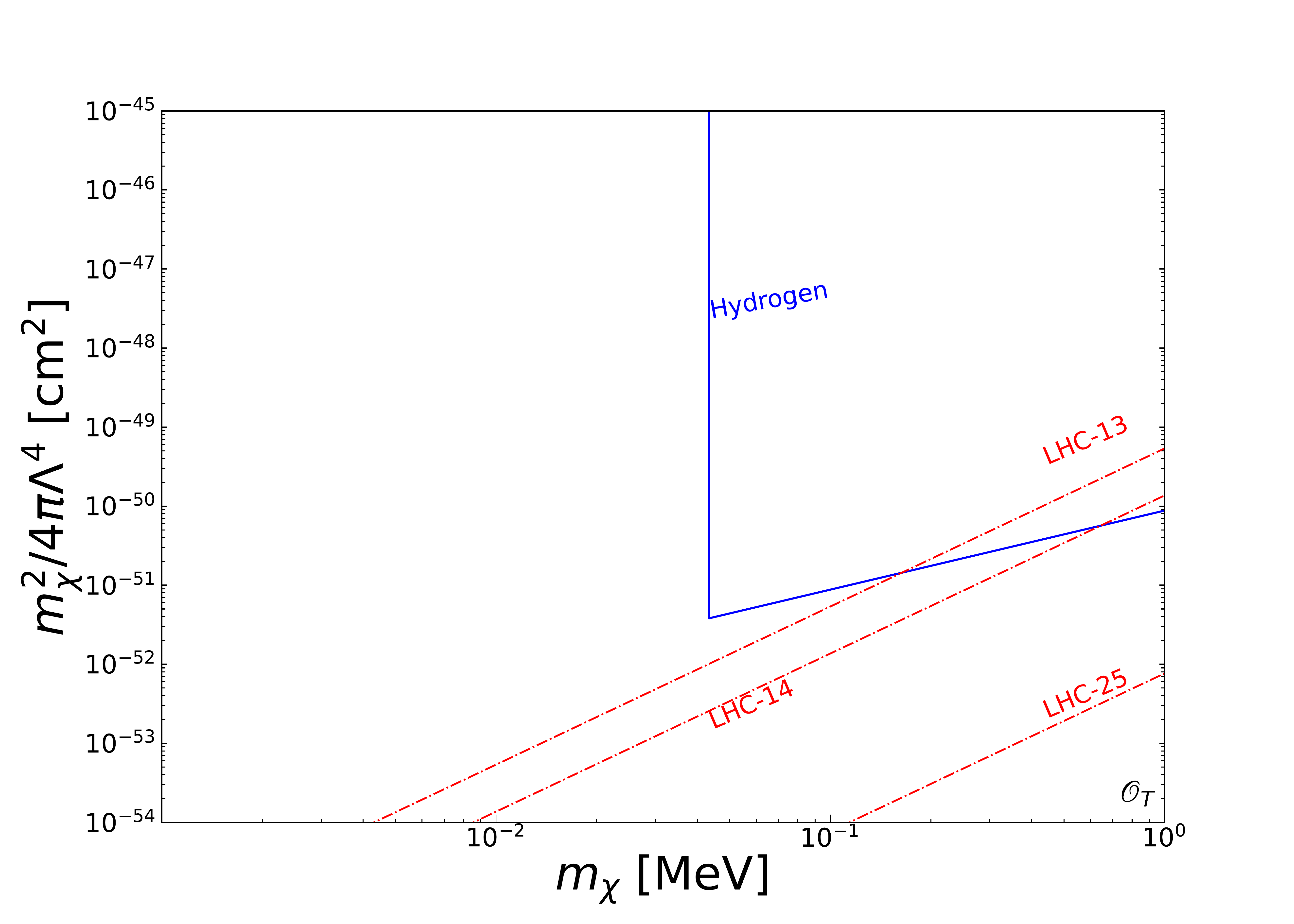}
\caption{\it
Excluded regions in the $m_\chi^2/4\pi\varLambda^4$-$m_\chi$ plane due to SD scatterings for the tensor operator.
The \textbf{left panel} represents the heavy target,
while the \textbf{right panel} is for the hydrogen target.
}
\label{fig:SD:T}
\end{figure}

\section{Conclusion}
\label{sec:conclusion}
In this paper, we investigate four-fermion contact operators that involve a dark fermion, a neutral neutrino, and a quark pair. We assess the sensitivity of these operators at the LHC through mono-lepton production processes. To validate our results, we compare the simulated distributions with the data from the ATLAS experiment.
We also emphasize the substantial potential of future upgrades to the LHC for exploring fermionic DM.
Moreover, we contrast the constraints from DM direct detection experiments with those obtained from LHC measurements. This comparison reveals the LHC's distinct advantage in probing light DM ($m_\chi \in [\sim 0, \sim 1]\mev$)
and at the TeV scale.

Nevertheless, direct detection experiments, such as Super-Kamiokande and Borexino, still provide competitive constraints. These experiments benefit from the use of sufficiently light nuclei and large exposure, along with a unique recoil energy spectrum.
In addition, for the operators $\mathcal{O}_S$ and $\mathcal{O}_V$, SI scattering with heavy nuclear targets shows well defined exclusion regions. In contrast, SD scattering for the operators $\mathcal{O}_P$, $\mathcal{O}_A$, and $\mathcal{O}_T$ exhibits more stringent constraints.

\appendix
\section{ Experimental Parameters for SI and SD Absorption} \label{App:A}
The Tab.~\ref{tab:SI:Exposure} presents the experiments analyzed in this paper for probing the SI absorption signal. Additionally, some typical experiments, which include isotopes with non-zero total spin, are listed in Tab.~\ref{tab:SD:Exposure}. Furthermore, we list the  spin expectation value $\mathbb{S}_N$  of some isotopes that will be studied
in this paper.

\begin{table}[th]
\renewcommand\arraystretch{1.2}
\begin{center}
\begin{tabular}{c c c c c }
Experiment  & Target & Exposure  & $E_{\rm R}^{\rm th}$
\\\hline\hline
Super-Kamiokande \cite{Super-Kamiokande:2019hga}
&
$\mathrm{H}_2\mathrm{O}$
&
171000 t yr
&
3.5 \mev
\\[1mm]\hline
CRESST-II \cite{CRESST:2015txj}
&
$\mathrm{Ca}\mathrm{W}\mathrm{O}_4$
&
$52$ kg day
&
307 eV
\\[1mm]\hline
CRESST-III  \cite{CRESST:2017cdd}
%\cite{CRESST:2017cdd}
&
$\mathrm{Ca}\mathrm{W}\mathrm{O}_4$
&

$2.39$ kg day
&
100 eV
\\[1mm]\hline
DarkSide-50 \cite{DarkSide:2014llq,DarkSide:2018bpj}
&
Liquid $\mathrm{Ar}$
&
$6787$ kg day
&
0.6 keV
\\[1mm]\hline
XENONnT \cite{XENON:2023cxc}
&
Liquid Xe
&
1.09 t yr
&
3 keV
\\[1mm]\hline
PandaX-4T \cite{PandaX:2023xgl} %\\  %\cite{PandaX:2022ood} for electron target
&
Liquid Xe
&
0.55 t yr
&
3 keV
\\[1mm]\hline
Borexino \cite{Borexino:2018pev}
&
$\mathrm{C}_6 \mathrm{H}_3\left(\mathrm{CH}_3\right)_3$
&
%817 t yr
958.58 t yr
% 596.21 t yr
&
500 keV
\\[1mm]\hline
PICO-60  \cite{PICO:2019vsc}
&
$\mathrm{CF}_3 \mathrm{I}$
&
3415 kg day
&
20 keV
%
%\\\hline\cline{1-3}\hline\hline\cline{1-3}\hline
%
\\[1mm]\hline
EDELWEISS-II \cite{EDELWEISS:2011epn}
&
Ge
&
384 kg day
&
20 keV
\\[1mm]\hline
COSINE-200 \cite{COSINE:2020egt}
&
NaI
&
0.6 t yr
&
1 keV
\\[1mm]\hline
LZ \cite{LZ:2024psa}
&
Liquid Xe
&
0.9 t yr
&
1 keV
\\[1mm]\hline
SuperCDMS  \cite{SuperCDMS:2022kgp}
&
Ge
&
36.9 kg day
&
70 eV
\\[1mm]\hline
DAMIC  \cite{SENSEI:2023rcc}
%\cite{PICO:2019vsc}
&
Si
&
3.25 kg day
&
23 eV
\\[1mm]\hline
DEAP-3600 \cite{Viel:2024qjs}
&
Liquid Ar
&
7.35 t yr
&
17.5 keV
\\[1mm]\hline
CDEX-50 \cite{CDEX:2023vvc}
&
Ge
&
150 jg yr
&
160 eV
\\[1mm]
\hline\hline
\end{tabular}
\caption{\it
Experiments studied here for probing the SI absorption signals.
}
\label{tab:SI:Exposure}
\end{center}
\end{table}

\begin{table}[th]
\renewcommand\arraystretch{1.2}
\begin{center}
\begin{tabular}{c c c c c c}
Experiment  & Target & Exposure & Isotope (Abund.) & $E_{\rm R}^{\rm th}$
\\\hline\hline
Super-Kamiokande \cite{Super-Kamiokande:2019hga}
&
$\mathrm{H}_2\mathrm{O}$
&
171000 t yr
&
$ {}^{1}_{1}\mathrm{H} $ (99.985\%)
&
3.5 \mev
\\[1mm]\hline
XENONnT \cite{XENON:2023cxc}
&
Liquid Xe
&
1.09 t yr
&
\begin{tabular}{c}
$ {}^{129}_{\;\;54}\mathrm{Xe} $ (26.4\%) \\
$ {}^{131}_{\;\;54}\mathrm{Xe} $ (21.2\%)
\end{tabular}
&
3 keV
\\[1mm]\hline
PandaX-4T \cite{PandaX:2022xas}
&
Liquid Xe
&
0.63 t yr
&
\begin{tabular}{c}
$ {}^{129}_{\;\;54}\mathrm{Xe} $ (26.4\%) \\
$ {}^{131}_{\;\;54}\mathrm{Xe} $ (21.2\%)
\end{tabular}
&
3 keV
\\[1mm]\hline
Borexino \cite{Borexino:2018pev}
&
$\mathrm{C}_6 \mathrm{H}_3\left(\mathrm{CH}_3\right)_3$
&
817 t yr
&
\begin{tabular}{c}
$ {}^{13}_{\;\,6}\mathrm{C} $ (1.1\%) \\
$ {}^{1}_{1}\mathrm{H} $ (99.985\%)
\end{tabular}
&
500 keV
\\[1mm]\hline
CRESST-III  \cite{CRESST:2022dtl}
&
$\mathrm{Li}\mathrm{Al}\mathrm{O}_2$
&
2.345 kg day
&
\begin{tabular}{c}
$ {}^{6}_{3}\mathrm{Li} $ (7.5\%) \\ $ {}^{7}_{3}\mathrm{Li} $ (92.5\%)
\\ $ {}^{27}_{13}\mathrm{Al} $ (100\%)
\end{tabular}
&
94.1 eV
\\[1mm]\hline
PICO-60 \cite{PICO:2019vsc}
&
$\mathrm{CF}_3 \mathrm{I}$
&
2207 kg day
&
\begin{tabular}{c}
$ {}^{13}_{\;\,6}\mathrm{C} $ (1.1\%) \\
$ {}^{19}_{\;\,9}\mathrm{F} $ (100\%)
\end{tabular}
&
3.3 keV
\\[1mm]\hline
PICASSO \cite{PICASSO:2012ngj}
&
F
&
114 kg day
&
\begin{tabular}{c}
$ {}^{19}_{\;\;9}\mathrm{F} $ (100\%)
\end{tabular}
&
1.7 keV
\\[1mm]\hline
COSINE-200 \cite{COSINE:2020egt}
&
NaI
&
0.6 t yr
&
\begin{tabular}{c}
$ {}^{23}_{\;\;11}\mathrm{Na} $ (100\%) \\
$ {}^{127}_{\;\;53}\mathrm{I} $ (100\%)
\end{tabular}
&
1 keV
\\[1mm]\hline
LZ \cite{LZ:2024psa}
&
Liquid Xe
&
0.9 t yr
&
\begin{tabular}{c}
$ {}^{129}_{\;\;54}\mathrm{Xe} $ (26.4\%) \\
$ {}^{131}_{\;\;54}\mathrm{Xe} $ (21.2\%)
\end{tabular}
&
1 keV
\\[1mm]\hline
SuperCDMS  \cite{SuperCDMS:2022kgp}
&
Ge
&
36.9 kg day
&
\begin{tabular}{c}
$ {}^{73}_{22}\mathrm{Ge} $ (27.31\%)
\end{tabular}
&
70 eV
\\[1mm]\hline
DAMIC  \cite{SENSEI:2023rcc}
&
Si
&
3.25 kg day
&
\begin{tabular}{c}
$ {}^{29}_{\;\,14}\mathrm{Si} $ (4.7\%)
\end{tabular}
&
23 eV
\\[1mm]
\hline\hline
\end{tabular}
\caption{\it
Experiments studied here for probing the SD absorption signals.
}
\label{tab:SD:Exposure}
\end{center}
\end{table}

\begin{table}[!h]
\renewcommand\arraystretch{1.2}
\begin{center}
\begin{tabular}{c c c c c c}
Isotope (Abund.) & $J$ & $\mathbb{S}_{p}$ & $\mathbb{S}_{n}$ & Ref.
\\\hline\hline
$ {}^{1}_{1}\mathrm{H} $ (99.985\%)  & 1/2 & 0.5 & 0 & \cite{Ellis:1987sh}
\\[1mm]\hline
$ {}^{6}_{3}\mathrm{Li} $ (7.5\%)  & 1/2 & 0.472 & 0.472 & \cite{Girlanda:2011fh,CRESST:2022dtl}
\\
$ {}^{7}_{3}\mathrm{Li} $ (92.5\%) & 3/2 & 0.497 & 0.004 & \cite{Pacheco:1989jz}
\\[1mm]\hline
$ {}^{13}_{\,\,\,6}\mathrm{C} $ (1.1\%)  & 1/2 & -0.009 & -0.172 & \cite{Engel:1989ix}
\\[1mm]\hline
$ {}^{27}_{13}\mathrm{Al} $ (100\%)   & 5/2 & 0.343 & 0.0296 & \cite{Engel:1995gw}
\\[1mm]\hline
$ {}^{73}_{22}\mathrm{Ge} $ (27.31\%)   & 9/2 & 0.37 & 0.123 & \cite{Stadnik:2014xja}
\\[1mm]\hline
$ {}^{23}_{11}\mathrm{Na} $ (100\%)  & 3/2 & -0.051 & -0.051 & \cite{Stadnik:2014xja}
\\[1mm]\hline
$ {}^{29}_{\,\,\,14}\mathrm{Si} $ (4.7\%)  & 1/2 & -0.009 & -0.172 & \cite{Stadnik:2014xja}
\\[1mm]\hline
$ {}^{19}_{\,\,\,9}\mathrm{F} $ (100\%)  & 1/2 & 0.339 & 0.161 & \cite{Divari:2000dc}
\\[1mm]\hline
$ {}^{127}_{53}\mathrm{I} $ (100\%)   & 5/2 & 0.265 & 0.235 & \cite{Stadnik:2014xja}
\\[1mm]\hline
\begin{tabular}{c}
$ {}^{129}_{\;\;54}\mathrm{Xe} $ (26.4\%) \\
$ {}^{131}_{\;\;54}\mathrm{Xe} $ (21.2\%)
\end{tabular}
&
\begin{tabular}{r} 1/2 \\  3/2  \end{tabular}
&
\begin{tabular}{r} 0.0128 \\  -0.012  \end{tabular}
&
\begin{tabular}{r} 0.300 \\   -0.217 \end{tabular}
&
\cite{Ressell:1997kx}
\\[1mm]
\hline\hline
\end{tabular}
\caption{\it
Spin expectation values ($\mathbb{S}_N$) of the isotopes studied in this paper.
}
\label{tab:exps}
\end{center}
\end{table}

\acknowledgments
K.M. was supported by the Natural Science Basic Research Program of Shaanxi (Program No. 2023-JC-YB-041).

\bibliographystyle{Nudamol}
\bibliography{Nudamol}

\providecommand{\href}[2]{#2}\begingroup\raggedright\begin{thebibliography}{100}

\bibitem{Navarro:1995iw}
J.F.~Navarro, C.S.~Frenk and S.D.M.~White, \emph{{The Structure of cold dark
  matter halos}}, \href{https://doi.org/10.1086/177173}{\emph{Astrophys. J.}
  {\bfseries 462} (1996) 563}
  [\href{https://arxiv.org/abs/astro-ph/9508025}{{\ttfamily
  astro-ph/9508025}}].

\bibitem{Clowe:2006eq}
D.~Clowe, M.~Bradac, A.H.~Gonzalez, M.~Markevitch, S.W.~Randall, C.~Jones
  et~al., \emph{{A direct empirical proof of the existence of dark matter}},
  \href{https://doi.org/10.1086/508162}{\emph{Astrophys. J. Lett.} {\bfseries
  648} (2006) L109} [\href{https://arxiv.org/abs/astro-ph/0608407}{{\ttfamily
  astro-ph/0608407}}].

\bibitem{Spergel:1999mh}
D.N.~Spergel and P.J.~Steinhardt, \emph{{Observational evidence for
  selfinteracting cold dark matter}},
  \href{https://doi.org/10.1103/PhysRevLett.84.3760}{\emph{Phys. Rev. Lett.}
  {\bfseries 84} (2000) 3760}
  [\href{https://arxiv.org/abs/astro-ph/9909386}{{\ttfamily
  astro-ph/9909386}}].

\bibitem{Bertone:2004pz}
G.~Bertone, D.~Hooper and J.~Silk, \emph{{Particle dark matter: Evidence,
  candidates and constraints}},
  \href{https://doi.org/10.1016/j.physrep.2004.08.031}{\emph{Phys. Rept.}
  {\bfseries 405} (2005) 279}
  [\href{https://arxiv.org/abs/hep-ph/0404175}{{\ttfamily hep-ph/0404175}}].

\bibitem{Boveia:2018yeb}
A.~Boveia and C.~Doglioni, \emph{{Dark Matter Searches at Colliders}},
  \href{https://doi.org/10.1146/annurev-nucl-101917-021008}{\emph{Ann. Rev.
  Nucl. Part. Sci.} {\bfseries 68} (2018) 429}
  [\href{https://arxiv.org/abs/1810.12238}{{\ttfamily 1810.12238}}].

\bibitem{Arcadi:2017atc}
G.~Arcadi, M.~Lindner, Y.~Mambrini, M.~Pierre and F.S.~Queiroz, \emph{{GUT
  Models at Current and Future Hadron Colliders and Implications to Dark Matter
  Searches}}, \href{https://doi.org/10.1016/j.physletb.2017.05.023}{\emph{Phys.
  Lett. B} {\bfseries 771} (2017) 508}
  [\href{https://arxiv.org/abs/1704.02328}{{\ttfamily 1704.02328}}].

\bibitem{Landsberg:2015kwa}
G.~Landsberg, \emph{{Moving Beyond Effective Field Theory in Dark Matter
  Searches at Colliders}},  in \emph{{50th Rencontres de Moriond on QCD and
  High Energy Interactions}}, pp.~211--216, 6, 2015
  [\href{https://arxiv.org/abs/1506.00660}{{\ttfamily 1506.00660}}].

\bibitem{Steigman:2012nb}
G.~Steigman, B.~Dasgupta and J.F.~Beacom, \emph{{Precise Relic WIMP Abundance
  and its Impact on Searches for Dark Matter Annihilation}},
  \href{https://doi.org/10.1103/PhysRevD.86.023506}{\emph{Phys. Rev. D}
  {\bfseries 86} (2012) 023506}
  [\href{https://arxiv.org/abs/1204.3622}{{\ttfamily 1204.3622}}].

\bibitem{deDiosZornoza:2021rgw}
J.~de~Dios~Zornoza, \emph{{Review on Indirect Dark Matter Searches with
  Neutrino Telescopes}},
  \href{https://doi.org/10.3390/universe7110415}{\emph{Universe} {\bfseries 7}
  (2021) 415}.

\bibitem{PerezdelosHeros:2020qyt}
C.~P\'erez de~los Heros, \emph{{Status, Challenges and Directions in Indirect
  Dark Matter Searches}},
  \href{https://doi.org/10.3390/sym12101648}{\emph{Symmetry} {\bfseries 12}
  (2020) 1648} [\href{https://arxiv.org/abs/2008.11561}{{\ttfamily
  2008.11561}}].

\bibitem{Cox:2023cjw}
P.~Cox, M.J.~Dolan and J.~Wood, \emph{{Absorption of Fermionic Dark Matter via
  the Scalar Portal}},  \href{https://arxiv.org/abs/2308.00309}{{\ttfamily
  2308.00309}}.

\bibitem{Ge:2024euk}
S.-F.~Ge and O.~Titov, \emph{{Incoherent Fermionic Dark Matter Absorption with
  Nucleon Fermi Motion}},  \href{https://arxiv.org/abs/2405.05728}{{\ttfamily
  2405.05728}}.

\bibitem{Ma:2024tkt}
K.~Ma, S.-F.~Ge, L.-Y.~He and N.~Zhou, \emph{{Complementary Search of Fermionic
  Absorption Operators at Hadron Collider and Direct Detection Experiments}},
  \href{https://arxiv.org/abs/2405.16878}{{\ttfamily 2405.16878}}.

\bibitem{Ma:2024aoc}
K.~Ma, \emph{{Exploring Four Fermion Contact Couplings of a Dark Fermion and an
  Electron at Hadron Colliders and Direct Detection Experiments}},
  \href{https://arxiv.org/abs/2404.06419}{{\ttfamily 2404.06419}}.

\bibitem{Hochberg:2016sqx}
Y.~Hochberg, T.~Lin and K.M.~Zurek, \emph{{Absorption of light dark matter in
  semiconductors}},
  \href{https://doi.org/10.1103/PhysRevD.95.023013}{\emph{Phys. Rev. D}
  {\bfseries 95} (2017) 023013}
  [\href{https://arxiv.org/abs/1608.01994}{{\ttfamily 1608.01994}}].

\bibitem{Hochberg:2016ajh}
Y.~Hochberg, T.~Lin and K.M.~Zurek, \emph{{Detecting Ultralight Bosonic Dark
  Matter via Absorption in Superconductors}},
  \href{https://doi.org/10.1103/PhysRevD.94.015019}{\emph{Phys. Rev. D}
  {\bfseries 94} (2016) 015019}
  [\href{https://arxiv.org/abs/1604.06800}{{\ttfamily 1604.06800}}].

\bibitem{Mitridate:2023izi}
A.~Mitridate, K.~Pardo, T.~Trickle and K.M.~Zurek, \emph{{Effective field
  theory for dark matter absorption on single phonons}},
  \href{https://doi.org/10.1103/PhysRevD.109.015010}{\emph{Phys. Rev. D}
  {\bfseries 109} (2024) 015010}
  [\href{https://arxiv.org/abs/2308.06314}{{\ttfamily 2308.06314}}].

\bibitem{Ge:2022ius}
S.-F.~Ge, X.-G.~He, X.-D.~Ma and J.~Sheng, \emph{{Revisiting the fermionic dark
  matter absorption on electron target}},
  \href{https://doi.org/10.1007/JHEP05(2022)191}{\emph{JHEP} {\bfseries 05}
  (2022) 191} [\href{https://arxiv.org/abs/2201.11497}{{\ttfamily
  2201.11497}}].

\bibitem{PandaX:2022ood}
{\scshape PandaX} collaboration, \emph{{Search for Light Fermionic Dark Matter
  Absorption on Electrons in PandaX-4T}},
  \href{https://doi.org/10.1103/PhysRevLett.129.161804}{\emph{Phys. Rev. Lett.}
  {\bfseries 129} (2022) 161804}
  [\href{https://arxiv.org/abs/2206.02339}{{\ttfamily 2206.02339}}].

\bibitem{CDEX:2024bum}
{\scshape CDEX} collaboration, \emph{{First Search for Light Fermionic Dark
  Matter Absorption on Electrons Using Germanium Detector in CDEX-10
  Experiment}},  \href{https://arxiv.org/abs/2404.09793}{{\ttfamily
  2404.09793}}.

\bibitem{Dror:2020czw}
J.A.~Dror, G.~Elor, R.~McGehee and T.-T.~Yu, \emph{{Absorption of sub-MeV
  fermionic dark matter by electron targets}},
  \href{https://doi.org/10.1103/PhysRevD.103.035001}{\emph{Phys. Rev. D}
  {\bfseries 103} (2021) 035001}
  [\href{https://arxiv.org/abs/2011.01940}{{\ttfamily 2011.01940}}].

\bibitem{Ge:2023wye}
S.-F.~Ge, K.~Ma, X.-D.~Ma and J.~Sheng, \emph{{Associated production of
  neutrino and dark fermion at future lepton colliders}},
  \href{https://doi.org/10.1007/JHEP11(2023)190}{\emph{JHEP} {\bfseries 11}
  (2023) 190} [\href{https://arxiv.org/abs/2306.00657}{{\ttfamily
  2306.00657}}].

\bibitem{Bai:2012xg}
Y.~Bai and T.M.P.~Tait, \emph{{Searches with Mono-Leptons}},
  \href{https://doi.org/10.1016/j.physletb.2013.05.057}{\emph{Phys. Lett. B}
  {\bfseries 723} (2013) 384}
  [\href{https://arxiv.org/abs/1208.4361}{{\ttfamily 1208.4361}}].

\bibitem{CMS:2013iea}
{\scshape CMS} collaboration, \emph{{Search for dark matter in the mono-lepton
  channel with pp collision events at center-of-mass energy of 8 TeV}}, .

\bibitem{Matsumoto:2018ioi}
S.~Matsumoto, S.~Shirai and M.~Takeuchi, \emph{{Indirect Probe of
  Electroweak-Interacting Particles with Mono-Lepton Signatures at Hadron
  Colliders}}, \href{https://doi.org/10.1007/JHEP03(2019)076}{\emph{JHEP}
  {\bfseries 03} (2019) 076}
  [\href{https://arxiv.org/abs/1810.12234}{{\ttfamily 1810.12234}}].

\bibitem{Dreiner:2013vla}
H.~Dreiner, D.~Schmeier and J.~Tattersall, \emph{{Contact Interactions Probe
  Effective Dark Matter Models at the LHC}},
  \href{https://doi.org/10.1209/0295-5075/102/51001}{\emph{EPL} {\bfseries 102}
  (2013) 51001} [\href{https://arxiv.org/abs/1303.3348}{{\ttfamily
  1303.3348}}].

\bibitem{Bishara:2017pfq}
F.~Bishara, J.~Brod, B.~Grinstein and J.~Zupan, \emph{{From quarks to nucleons
  in dark matter direct detection}},
  \href{https://doi.org/10.1007/JHEP11(2017)059}{\emph{JHEP} {\bfseries 11}
  (2017) 059} [\href{https://arxiv.org/abs/1707.06998}{{\ttfamily
  1707.06998}}].

\bibitem{Belyaev:2018pqr}
A.~Belyaev, E.~Bertuzzo, C.~Caniu~Barros, O.~Eboli, G.~Grilli Di~Cortona,
  F.~Iocco et~al., \emph{{Interplay of the LHC and non-LHC Dark Matter searches
  in the Effective Field Theory approach}},
  \href{https://doi.org/10.1103/PhysRevD.99.015006}{\emph{Phys. Rev. D}
  {\bfseries 99} (2019) 015006}
  [\href{https://arxiv.org/abs/1807.03817}{{\ttfamily 1807.03817}}].

\bibitem{Ma:2024gqj}
K.~Ma and L.-Y.~He, \emph{{Constraining Gluonic Contact Interaction of a
  Neutrino-philic Dark Fermion at Hadron Colliders and Direct Detection
  Experiments}},  \href{https://arxiv.org/abs/2405.20890}{{\ttfamily
  2405.20890}}.

\bibitem{Abdallah:2015uba}
W.~Abdallah, J.~Fiaschi, S.~Khalil and S.~Moretti, \emph{{Mono-jet, -photon and
  -Z signals of a supersymmetric (B \ensuremath{-} L) model at the Large Hadron
  Collider}}, \href{https://doi.org/10.1007/JHEP02(2016)157}{\emph{JHEP}
  {\bfseries 02} (2016) 157}
  [\href{https://arxiv.org/abs/1510.06475}{{\ttfamily 1510.06475}}].

\bibitem{Bai:2015nfa}
Y.~Bai, J.~Bourbeau and T.~Lin, \emph{{Dark matter searches with a
  mono-$Z^{\prime}$ jet}},
  \href{https://doi.org/10.1007/JHEP06(2015)205}{\emph{JHEP} {\bfseries 06}
  (2015) 205} [\href{https://arxiv.org/abs/1504.01395}{{\ttfamily
  1504.01395}}].

\bibitem{daSilveira:2023hmt}
G.G.~da~Silveira and M.S.~Mateus, \emph{{Investigation of spin-dependent dark
  matter in mono-photon production at high-energy colliders}},
  \href{https://arxiv.org/abs/2308.03680}{{\ttfamily 2308.03680}}.

\bibitem{Kalinowski:2022unu}
J.~Kalinowski, W.~Kotlarski, K.~Mekala, A.F.~Zarnecki and K.~Zembaczynski,
  \emph{{New approach to DM searches with mono-photon signature}},
  \href{https://doi.org/10.22323/1.414.0294}{\emph{PoS} {\bfseries ICHEP2022}
  (2022) 294}.

\bibitem{Gershtein:2008bf}
Y.~Gershtein, F.~Petriello, S.~Quackenbush and K.M.~Zurek, \emph{{Discovering
  hidden sectors with mono-photon $Z^\prime$o searches}},
  \href{https://doi.org/10.1103/PhysRevD.78.095002}{\emph{Phys. Rev. D}
  {\bfseries 78} (2008) 095002}
  [\href{https://arxiv.org/abs/0809.2849}{{\ttfamily 0809.2849}}].

\bibitem{H:2024mff}
A.M.H.~H., E.-s.A.~El-dahshan and S.~Elgammal, \emph{{Investigating a new
  neutral heavy gauge boson within the mono-Z$^{\prime}$ model via simulated pp
  collisions at $\sqrt{s}$ = 14 TeV at the HL-LHC}},
  \href{https://arxiv.org/abs/2406.19004}{{\ttfamily 2406.19004}}.

\bibitem{Benitez-Irarrazabal:2024ich}
G.~Ben\'\i{}tez-Irarr\'azabal and A.R.~Zerwekh, \emph{{Mono-Higgs and Mono-Z
  Production in the Minimal Vector Dark Matter Model}},
  \href{https://doi.org/10.3390/universe10070288}{\emph{Universe} {\bfseries
  10} (2024) 288} [\href{https://arxiv.org/abs/2401.03954}{{\ttfamily
  2401.03954}}].

\bibitem{Kundu:2021cmo}
S.~Kundu, A.~Guha, P.K.~Das and P.S.B.~Dev, \emph{{EFT analysis of leptophilic
  dark matter at future electron-positron colliders in the mono-photon and
  mono-Z channels}},
  \href{https://doi.org/10.1103/PhysRevD.107.015003}{\emph{Phys. Rev. D}
  {\bfseries 107} (2023) 015003}
  [\href{https://arxiv.org/abs/2110.06903}{{\ttfamily 2110.06903}}].

\bibitem{Wan:2018eaz}
N.~Wan, N.~Li, B.~Zhang, H.~Yang, M.-F.~Zhao, M.~Song et~al., \emph{{Searches
  for Dark Matter via Mono-W Production in Inert Doublet Model at the LHC}},
  \href{https://doi.org/10.1088/0253-6102/69/5/617}{\emph{Commun. Theor. Phys.}
  {\bfseries 69} (2018) 617}.

\bibitem{Bell:2015rdw}
N.F.~Bell, Y.~Cai and R.K.~Leane, \emph{{Mono-W Dark Matter Signals at the LHC:
  Simplified Model Analysis}},
  \href{https://doi.org/10.1088/1475-7516/2016/01/051}{\emph{JCAP} {\bfseries
  01} (2016) 051} [\href{https://arxiv.org/abs/1512.00476}{{\ttfamily
  1512.00476}}].

\bibitem{Acharya:2024vsu}
{\scshape CMS} collaboration, \emph{{Search for Dark Matter with mono-X
  Signatures in CMS}}, \href{https://doi.org/10.22323/1.441.0074}{\emph{PoS}
  {\bfseries TAUP2023} (2024) 074}.

\bibitem{Bhattacharya:2022qck}
S.~Bhattacharya, P.~Ghosh, J.~Lahiri and B.~Mukhopadhyaya, \emph{{Mono-X signal
  and two component dark matter: new distinction criteria}},
  \href{https://arxiv.org/abs/2211.10749}{{\ttfamily 2211.10749}}.

\bibitem{Bernreuther:2018nat}
E.~Bernreuther, J.~Horak, T.~Plehn and A.~Butter, \emph{{Actual Physics behind
  Mono-X}}, \href{https://doi.org/10.21468/SciPostPhys.5.4.034}{\emph{SciPost
  Phys.} {\bfseries 5} (2018) 034}
  [\href{https://arxiv.org/abs/1805.11637}{{\ttfamily 1805.11637}}].

\bibitem{PandaX:2022osq}
{\scshape PandaX} collaboration, \emph{{First Search for the Absorption of
  Fermionic Dark Matter with the PandaX-4T Experiment}},
  \href{https://doi.org/10.1103/PhysRevLett.129.161803}{\emph{Phys. Rev. Lett.}
  {\bfseries 129} (2022) 161803}
  [\href{https://arxiv.org/abs/2205.15771}{{\ttfamily 2205.15771}}].

\bibitem{Dror:2019dib}
J.A.~Dror, G.~Elor and R.~Mcgehee, \emph{{Absorption of Fermionic Dark Matter
  by Nuclear Targets}},
  \href{https://doi.org/10.1007/JHEP02(2020)134}{\emph{JHEP} {\bfseries 02}
  (2020) 134} [\href{https://arxiv.org/abs/1908.10861}{{\ttfamily
  1908.10861}}].

\bibitem{Li:2022kca}
T.~Li, J.~Liao and R.-J.~Zhang, \emph{{Dark magnetic dipole property in
  fermionic absorption by nucleus and electrons}},
  \href{https://doi.org/10.1007/JHEP05(2022)071}{\emph{JHEP} {\bfseries 05}
  (2022) 071} [\href{https://arxiv.org/abs/2201.11905}{{\ttfamily
  2201.11905}}].

\bibitem{Dror:2019onn}
J.A.~Dror, G.~Elor and R.~Mcgehee, \emph{{Directly Detecting Signals from
  Absorption of Fermionic Dark Matter}},
  \href{https://doi.org/10.1103/PhysRevLett.124.181301}{\emph{Phys. Rev. Lett.}
  {\bfseries 124} (2020) 18}
  [\href{https://arxiv.org/abs/1905.12635}{{\ttfamily 1905.12635}}].

\bibitem{Belfatto:2021ats}
B.~Belfatto, D.~Buttazzo, C.~Gross, P.~Panci, A.~Strumia, N.~Vignaroli et~al.,
  \emph{{Dark Matter abundance via thermal decays and leptoquark mediators}},
  \href{https://doi.org/10.1007/JHEP06(2022)084}{\emph{JHEP} {\bfseries 06}
  (2022) 084} [\href{https://arxiv.org/abs/2111.14808}{{\ttfamily
  2111.14808}}].

\bibitem{Liu:2016idz}
D.~Liu, A.~Pomarol, R.~Rattazzi and F.~Riva, \emph{{Patterns of Strong Coupling
  for LHC Searches}},
  \href{https://doi.org/10.1007/JHEP11(2016)141}{\emph{JHEP} {\bfseries 11}
  (2016) 141} [\href{https://arxiv.org/abs/1603.03064}{{\ttfamily
  1603.03064}}].

\bibitem{Giudice:2007fh}
G.F.~Giudice, C.~Grojean, A.~Pomarol and R.~Rattazzi, \emph{{The
  Strongly-Interacting Light Higgs}},
  \href{https://doi.org/10.1088/1126-6708/2007/06/045}{\emph{JHEP} {\bfseries
  06} (2007) 045} [\href{https://arxiv.org/abs/hep-ph/0703164}{{\ttfamily
  hep-ph/0703164}}].

\bibitem{Busoni:2013lha}
G.~Busoni, A.~De~Simone, E.~Morgante and A.~Riotto, \emph{{On the Validity of
  the Effective Field Theory for Dark Matter Searches at the LHC}},
  \href{https://doi.org/10.1016/j.physletb.2013.11.069}{\emph{Phys. Lett. B}
  {\bfseries 728} (2014) 412}
  [\href{https://arxiv.org/abs/1307.2253}{{\ttfamily 1307.2253}}].

\bibitem{Busoni:2014sya}
G.~Busoni, A.~De~Simone, J.~Gramling, E.~Morgante and A.~Riotto, \emph{{On the
  Validity of the Effective Field Theory for Dark Matter Searches at the LHC,
  Part II: Complete Analysis for the $s$-channel}},
  \href{https://doi.org/10.1088/1475-7516/2014/06/060}{\emph{JCAP} {\bfseries
  06} (2014) 060} [\href{https://arxiv.org/abs/1402.1275}{{\ttfamily
  1402.1275}}].

\bibitem{Busoni:2014haa}
G.~Busoni, A.~De~Simone, T.~Jacques, E.~Morgante and A.~Riotto, \emph{{On the
  Validity of the Effective Field Theory for Dark Matter Searches at the LHC
  Part III: Analysis for the $t$-channel}},
  \href{https://doi.org/10.1088/1475-7516/2014/09/022}{\emph{JCAP} {\bfseries
  09} (2014) 022} [\href{https://arxiv.org/abs/1405.3101}{{\ttfamily
  1405.3101}}].

\bibitem{Hill:2011be}
R.J.~Hill and M.P.~Solon, \emph{{Universal behavior in the scattering of heavy,
  weakly interacting dark matter on nuclear targets}},
  \href{https://doi.org/10.1016/j.physletb.2012.01.013}{\emph{Phys. Lett. B}
  {\bfseries 707} (2012) 539}
  [\href{https://arxiv.org/abs/1111.0016}{{\ttfamily 1111.0016}}].

\bibitem{Frandsen:2012db}
M.T.~Frandsen, U.~Haisch, F.~Kahlhoefer, P.~Mertsch and K.~Schmidt-Hoberg,
  \emph{{Loop-induced dark matter direct detection signals from gamma-ray
  lines}}, \href{https://doi.org/10.1088/1475-7516/2012/10/033}{\emph{JCAP}
  {\bfseries 10} (2012) 033} [\href{https://arxiv.org/abs/1207.3971}{{\ttfamily
  1207.3971}}].

\bibitem{Vecchi:2013iza}
L.~Vecchi, \emph{{WIMPs and Un-Naturalness}},
  \href{https://arxiv.org/abs/1312.5695}{{\ttfamily 1312.5695}}.

\bibitem{Crivellin:2014qxa}
A.~Crivellin, F.~D'Eramo and M.~Procura, \emph{{New Constraints on Dark Matter
  Effective Theories from Standard Model Loops}},
  \href{https://doi.org/10.1103/PhysRevLett.112.191304}{\emph{Phys. Rev. Lett.}
  {\bfseries 112} (2014) 191304}
  [\href{https://arxiv.org/abs/1402.1173}{{\ttfamily 1402.1173}}].

\bibitem{DEramo:2014nmf}
F.~D'Eramo and M.~Procura, \emph{{Connecting Dark Matter UV Complete Models to
  Direct Detection Rates via Effective Field Theory}},
  \href{https://doi.org/10.1007/JHEP04(2015)054}{\emph{JHEP} {\bfseries 04}
  (2015) 054} [\href{https://arxiv.org/abs/1411.3342}{{\ttfamily 1411.3342}}].

\bibitem{DEramo:2016gos}
F.~D'Eramo, B.J.~Kavanagh and P.~Panci, \emph{{You can hide but you have to
  run: direct detection with vector mediators}},
  \href{https://doi.org/10.1007/JHEP08(2016)111}{\emph{JHEP} {\bfseries 08}
  (2016) 111} [\href{https://arxiv.org/abs/1605.04917}{{\ttfamily
  1605.04917}}].

\bibitem{PIENU:2017wbj}
{\scshape PIENU} collaboration, \emph{{Improved search for heavy neutrinos in
  the decay $\pi\rightarrow e\nu$}},
  \href{https://doi.org/10.1103/PhysRevD.97.072012}{\emph{Phys. Rev. D}
  {\bfseries 97} (2018) 072012}
  [\href{https://arxiv.org/abs/1712.03275}{{\ttfamily 1712.03275}}].

\bibitem{Gori:2022vri}
S.~Gori et~al., \emph{{Dark Sector Physics at High-Intensity Experiments}},
  \href{https://arxiv.org/abs/2209.04671}{{\ttfamily 2209.04671}}.

\bibitem{Marra:2019lyc}
V.~Marra, R.~Rosenfeld and R.~Sturani, \emph{{Observing the dark sector}},
  \href{https://doi.org/10.3390/universe5060137}{\emph{Universe} {\bfseries 5}
  (2019) 137} [\href{https://arxiv.org/abs/1904.00774}{{\ttfamily
  1904.00774}}].

\bibitem{Deliyergiyev:2015oxa}
M.~Deliyergiyev, \emph{{Recent Progress in Search for Dark Sector Signatures}},
  \href{https://doi.org/10.1515/phys-2016-0034}{\emph{Open Phys.} {\bfseries
  14} (2016) 281} [\href{https://arxiv.org/abs/1510.06927}{{\ttfamily
  1510.06927}}].

\bibitem{Hofmann:2020wvr}
R.~Hofmann, \emph{{An SU(2) Gauge Principle for the Cosmic Microwave
  Background: Perspectives on the Dark Sector of the Cosmological Model}},
  \href{https://doi.org/10.3390/universe6090135}{\emph{Universe} {\bfseries 6}
  (2020) 135} [\href{https://arxiv.org/abs/2009.03734}{{\ttfamily
  2009.03734}}].

\bibitem{Lagouri:2022ier}
T.~Lagouri, \emph{{Review on Higgs Hidden\textendash{}Dark Sector Physics at
  High-Energy Colliders}},
  \href{https://doi.org/10.3390/sym14071299}{\emph{Symmetry} {\bfseries 14}
  (2022) 1299}.

\bibitem{ATLAS:2021kxv}
{\scshape ATLAS} collaboration, \emph{{Search for new phenomena in events with
  an energetic jet and missing transverse momentum in $pp$ collisions at $\sqrt
  {s}$ =13 TeV with the ATLAS detector}},
  \href{https://doi.org/10.1103/PhysRevD.103.112006}{\emph{Phys. Rev. D}
  {\bfseries 103} (2021) 112006}
  [\href{https://arxiv.org/abs/2102.10874}{{\ttfamily 2102.10874}}].

\bibitem{ATLAS:2019lsy}
{\scshape ATLAS} collaboration, \emph{{Search for a heavy charged boson in
  events with a charged lepton and missing transverse momentum from $pp$
  collisions at $\sqrt{s} = 13$ TeV with the ATLAS detector}},
  \href{https://doi.org/10.1103/PhysRevD.100.052013}{\emph{Phys. Rev. D}
  {\bfseries 100} (2019) 052013}
  [\href{https://arxiv.org/abs/1906.05609}{{\ttfamily 1906.05609}}].

\bibitem{ATLAS:2017jbq}
{\scshape ATLAS} collaboration, \emph{{Search for a new heavy gauge boson
  resonance decaying into a lepton and missing transverse momentum in 36
  fb$^{-1}$ of $pp$ collisions at $\sqrt{s} =$ 13 TeV with the ATLAS
  experiment}},
  \href{https://doi.org/10.1140/epjc/s10052-018-5877-y}{\emph{Eur. Phys. J. C}
  {\bfseries 78} (2018) 401}
  [\href{https://arxiv.org/abs/1706.04786}{{\ttfamily 1706.04786}}].

\bibitem{CMS:2018hff}
{\scshape CMS} collaboration, \emph{{Search for high-mass resonances in final
  states with a lepton and missing transverse momentum at $ \sqrt{s}=13 $
  TeV}}, \href{https://doi.org/10.1007/JHEP06(2018)128}{\emph{JHEP} {\bfseries
  06} (2018) 128} [\href{https://arxiv.org/abs/1803.11133}{{\ttfamily
  1803.11133}}].

\bibitem{CidVidal:2018eel}
X.~Cid~Vidal et~al., \emph{{Report from Working Group 3}: {Beyond the Standard
  Model physics at the HL-LHC and HE-LHC}},
  \href{https://doi.org/10.23731/CYRM-2019-007.585}{\emph{CERN Yellow Rep.
  Monogr.} {\bfseries 7} (2019) 585}
  [\href{https://arxiv.org/abs/1812.07831}{{\ttfamily 1812.07831}}].

\bibitem{CEPCStudyGroup:2023quu}
{\scshape CEPC Study Group} collaboration, \emph{{CEPC Technical Design Report:
  Accelerator}},
  \href{https://doi.org/10.1007/s41605-024-00463-y}{\emph{Radiat. Detect.
  Technol. Methods} {\bfseries 8} (2024) 1}
  [\href{https://arxiv.org/abs/2312.14363}{{\ttfamily 2312.14363}}].

\bibitem{CEPCStudyGroup:2018ghi}
{\scshape CEPC Study Group} collaboration, \emph{{CEPC Conceptual Design
  Report: Volume 2 - Physics \& Detector}},
  \href{https://arxiv.org/abs/1811.10545}{{\ttfamily 1811.10545}}.

\bibitem{Helm:1956zz}
R.H.~Helm, \emph{Inelastic and elastic scattering of 187-mev electrons from
  selected even-even nuclei},
  \href{https://doi.org/10.1103/PhysRev.104.1466}{\emph{Phys. Rev.} {\bfseries
  104} (1956) 1466}.

\bibitem{Lewin:1995rx}
J.D.~Lewin and P.F.~Smith, \emph{{Review of mathematics, numerical factors, and
  corrections for dark matter experiments based on elastic nuclear recoil}},
  \href{https://doi.org/10.1016/S0927-6505(96)00047-3}{\emph{Astropart. Phys.}
  {\bfseries 6} (1996) 87}.

\bibitem{Fitzpatrick:2012ix}
A.L.~Fitzpatrick, W.~Haxton, E.~Katz, N.~Lubbers and Y.~Xu, \emph{{The
  Effective Field Theory of Dark Matter Direct Detection}},
  \href{https://doi.org/10.1088/1475-7516/2013/02/004}{\emph{JCAP} {\bfseries
  02} (2013) 004} [\href{https://arxiv.org/abs/1203.3542}{{\ttfamily
  1203.3542}}].

\bibitem{Stadnik:2014xja}
Y.V.~Stadnik and V.V.~Flambaum, \emph{{Nuclear spin-dependent interactions:
  Searches for WIMP, Axion and Topological Defect Dark Matter, and Tests of
  Fundamental Symmetries}},
  \href{https://doi.org/10.1140/epjc/s10052-015-3326-8}{\emph{Eur. Phys. J. C}
  {\bfseries 75} (2015) 110} [\href{https://arxiv.org/abs/1408.2184}{{\ttfamily
  1408.2184}}].

\bibitem{Super-Kamiokande:2019hga}
{\scshape Super-Kamiokande} collaboration, \emph{{Measurement of the
  neutrino-oxygen neutral-current quasielastic cross section using atmospheric
  neutrinos at Super-Kamiokande}},
  \href{https://doi.org/10.1103/PhysRevD.99.032005}{\emph{Phys. Rev. D}
  {\bfseries 99} (2019) 032005}
  [\href{https://arxiv.org/abs/1901.05281}{{\ttfamily 1901.05281}}].

\bibitem{CRESST:2015txj}
{\scshape CRESST} collaboration, \emph{{Results on light dark matter particles
  with a low-threshold CRESST-II detector}},
  \href{https://doi.org/10.1140/epjc/s10052-016-3877-3}{\emph{Eur. Phys. J. C}
  {\bfseries 76} (2016) 25} [\href{https://arxiv.org/abs/1509.01515}{{\ttfamily
  1509.01515}}].

\bibitem{CRESST:2017cdd}
{\scshape CRESST} collaboration, \emph{{First results on low-mass dark matter
  from the CRESST-III experiment}},
  \href{https://doi.org/10.1088/1742-6596/1342/1/012076}{\emph{J. Phys. Conf.
  Ser.} {\bfseries 1342} (2020) 012076}
  [\href{https://arxiv.org/abs/1711.07692}{{\ttfamily 1711.07692}}].

\bibitem{DarkSide:2014llq}
{\scshape DarkSide} collaboration, \emph{{First Results from the DarkSide-50
  Dark Matter Experiment at Laboratori Nazionali del Gran Sasso}},
  \href{https://doi.org/10.1016/j.physletb.2015.03.012}{\emph{Phys. Lett. B}
  {\bfseries 743} (2015) 456}
  [\href{https://arxiv.org/abs/1410.0653}{{\ttfamily 1410.0653}}].

\bibitem{DarkSide:2018bpj}
{\scshape DarkSide} collaboration, \emph{{Low-Mass Dark Matter Search with the
  DarkSide-50 Experiment}},
  \href{https://doi.org/10.1103/PhysRevLett.121.081307}{\emph{Phys. Rev. Lett.}
  {\bfseries 121} (2018) 081307}
  [\href{https://arxiv.org/abs/1802.06994}{{\ttfamily 1802.06994}}].

\bibitem{XENON:2023cxc}
{\scshape XENON} collaboration, \emph{{First Dark Matter Search with Nuclear
  Recoils from the XENONnT Experiment}},
  \href{https://doi.org/10.1103/PhysRevLett.131.041003}{\emph{Phys. Rev. Lett.}
  {\bfseries 131} (2023) 041003}
  [\href{https://arxiv.org/abs/2303.14729}{{\ttfamily 2303.14729}}].

\bibitem{PandaX:2023xgl}
{\scshape PandaX} collaboration, \emph{{Search for
  Dark-Matter\textendash{}Nucleon Interactions with a Dark Mediator in
  PandaX-4T}},
  \href{https://doi.org/10.1103/PhysRevLett.131.191002}{\emph{Phys. Rev. Lett.}
  {\bfseries 131} (2023) 191002}
  [\href{https://arxiv.org/abs/2308.01540}{{\ttfamily 2308.01540}}].

\bibitem{Borexino:2018pev}
{\scshape Borexino} collaboration, \emph{{Modulations of the Cosmic Muon Signal
  in Ten Years of Borexino Data}},
  \href{https://doi.org/10.1088/1475-7516/2019/02/046}{\emph{JCAP} {\bfseries
  02} (2019) 046} [\href{https://arxiv.org/abs/1808.04207}{{\ttfamily
  1808.04207}}].

\bibitem{PICO:2019vsc}
{\scshape PICO} collaboration, \emph{{Dark Matter Search Results from the
  Complete Exposure of the PICO-60 C$_3$F$_8$ Bubble Chamber}},
  \href{https://doi.org/10.1103/PhysRevD.100.022001}{\emph{Phys. Rev. D}
  {\bfseries 100} (2019) 022001}
  [\href{https://arxiv.org/abs/1902.04031}{{\ttfamily 1902.04031}}].

\bibitem{EDELWEISS:2011epn}
{\scshape EDELWEISS} collaboration, \emph{{Final results of the EDELWEISS-II
  WIMP search using a 4-kg array of cryogenic germanium detectors with
  interleaved electrodes}},
  \href{https://doi.org/10.1016/j.physletb.2011.07.034}{\emph{Phys. Lett. B}
  {\bfseries 702} (2011) 329}
  [\href{https://arxiv.org/abs/1103.4070}{{\ttfamily 1103.4070}}].

\bibitem{COSINE:2020egt}
{\scshape COSINE} collaboration, \emph{{Development of ultra-pure NaI(Tl)
  detectors for the COSINE-200 experiment}},
  \href{https://doi.org/10.1140/epjc/s10052-020-8386-8}{\emph{Eur. Phys. J. C}
  {\bfseries 80} (2020) 814}
  [\href{https://arxiv.org/abs/2004.06287}{{\ttfamily 2004.06287}}].

\bibitem{LZ:2024psa}
{\scshape LZ} collaboration, \emph{{New constraints on ultraheavy dark matter
  from the LZ experiment}},
  \href{https://doi.org/10.1103/PhysRevD.109.112010}{\emph{Phys. Rev. D}
  {\bfseries 109} (2024) 112010}
  [\href{https://arxiv.org/abs/2402.08865}{{\ttfamily 2402.08865}}].

\bibitem{SuperCDMS:2022kgp}
{\scshape SuperCDMS} collaboration, \emph{{A Search for Low-mass Dark Matter
  via Bremsstrahlung Radiation and the Migdal Effect in SuperCDMS}},
  \href{https://arxiv.org/abs/2203.02594}{{\ttfamily 2203.02594}}.

\bibitem{SENSEI:2023rcc}
{\scshape SENSEI, DAMIC-M, DAMIC} collaboration, \emph{{Confirmation of the
  spectral excess in DAMIC at SNOLAB with skipper CCDs}},
  \href{https://doi.org/10.1103/PhysRevD.109.062007}{\emph{Phys. Rev. D}
  {\bfseries 109} (2024) 062007}
  [\href{https://arxiv.org/abs/2306.01717}{{\ttfamily 2306.01717}}].

\bibitem{Viel:2024qjs}
{\scshape DEAP-3600} collaboration, \emph{{Latest results from the DEAP-3600
  experiment at SNOLAB}}, \href{https://doi.org/10.22323/1.441.0075}{\emph{PoS}
  {\bfseries TAUP2023} (2024) 075}.

\bibitem{CDEX:2023vvc}
{\scshape CDEX} collaboration, \emph{{Projected WIMP sensitivity of the CDEX-50
  dark matter experiment}},
  \href{https://doi.org/10.1088/1475-7516/2024/07/009}{\emph{JCAP} {\bfseries
  07} (2024) 009} [\href{https://arxiv.org/abs/2309.01843}{{\ttfamily
  2309.01843}}].

\bibitem{PandaX:2022xas}
{\scshape PandaX} collaboration, \emph{{Constraints on the axial-vector and
  pseudo-scalar mediated WIMP-nucleus interactions from PandaX-4T experiment}},
  \href{https://doi.org/10.1016/j.physletb.2022.137487}{\emph{Phys. Lett. B}
  {\bfseries 834} (2022) 137487}
  [\href{https://arxiv.org/abs/2208.03626}{{\ttfamily 2208.03626}}].

\bibitem{CRESST:2022dtl}
{\scshape CRESST} collaboration, \emph{{Testing spin-dependent dark matter
  interactions with lithium aluminate targets in CRESST-III}},
  \href{https://doi.org/10.1103/PhysRevD.106.092008}{\emph{Phys. Rev. D}
  {\bfseries 106} (2022) 092008}
  [\href{https://arxiv.org/abs/2207.07640}{{\ttfamily 2207.07640}}].

\bibitem{PICASSO:2012ngj}
{\scshape PICASSO} collaboration, \emph{{Constraints on Low-Mass WIMP
  Interactions on $^{19}F$ from PICASSO}},
  \href{https://doi.org/10.1016/j.physletb.2012.03.078}{\emph{Phys. Lett. B}
  {\bfseries 711} (2012) 153}
  [\href{https://arxiv.org/abs/1202.1240}{{\ttfamily 1202.1240}}].

\bibitem{Ellis:1987sh}
J.R.~Ellis and R.A.~Flores, \emph{{Realistic predictions for the detection of
  supersymmetric dark matter}},
  \href{https://doi.org/10.1016/0550-3213(88)90111-3}{\emph{Nucl. Phys. B}
  {\bfseries 307} (1988) 883}.

\bibitem{Girlanda:2011fh}
L.~Girlanda, A.~Kievsky and M.~Viviani, \emph{{Subleading contributions to the
  three-nucleon contact interaction}},
  \href{https://doi.org/10.1103/PhysRevC.84.014001}{\emph{Phys. Rev. C}
  {\bfseries 84} (2011) 014001}
  [\href{https://arxiv.org/abs/1102.4799}{{\ttfamily 1102.4799}}].

\bibitem{Pacheco:1989jz}
A.F.~Pacheco and D.~Strottman, \emph{{Nuclear Structure Corrections to
  Estimates of the Spin Dependent {WIMP} Nucleus Cross-section}},
  \href{https://doi.org/10.1103/PhysRevD.40.2131}{\emph{Phys. Rev. D}
  {\bfseries 40} (1989) 2131}.

\bibitem{Engel:1989ix}
J.~Engel and P.~Vogel, \emph{{Spin dependent cross-sections of weakly
  interacting massive particles on nuclei}},
  \href{https://doi.org/10.1103/PhysRevD.40.3132}{\emph{Phys. Rev. D}
  {\bfseries 40} (1989) 3132}.

\bibitem{Engel:1995gw}
J.~Engel, M.T.~Ressell, I.S.~Towner and W.E.~Ormand, \emph{{Response of mica to
  weakly interacting massive particles}},
  \href{https://doi.org/10.1103/PhysRevC.52.2216}{\emph{Phys. Rev. C}
  {\bfseries 52} (1995) 2216}
  [\href{https://arxiv.org/abs/hep-ph/9504322}{{\ttfamily hep-ph/9504322}}].

\bibitem{Divari:2000dc}
P.C.~Divari, T.S.~Kosmas, J.D.~Vergados and L.D.~Skouras, \emph{{Shell model
  calculations for light supersymmetric particle scattering off light nuclei}},
  \href{https://doi.org/10.1103/PhysRevC.61.054612}{\emph{Phys. Rev. C}
  {\bfseries 61} (2000) 054612}.

\bibitem{Ressell:1997kx}
M.T.~Ressell and D.J.~Dean, \emph{{Spin dependent neutralino - nucleus
  scattering for A approximately 127 nuclei}},
  \href{https://doi.org/10.1103/PhysRevC.56.535}{\emph{Phys. Rev. C} {\bfseries
  56} (1997) 535} [\href{https://arxiv.org/abs/hep-ph/9702290}{{\ttfamily
  hep-ph/9702290}}].

\end{thebibliography}\endgroup

\end{document}